\def\Jl#1#2{#1 {\bf #2},\ }

\def\ApJ#1 {\Jl{Astroph. J.}{#1}}
\def\CQG#1 {\Jl{Class. Quantum Grav.}{#1}}
\def\DAN#1 {\Jl{Dokl. AN SSSR}{#1}}
\def\GC#1 {\Jl{Grav. Cosmol.}{#1}}
\def\GRG#1 {\Jl{Gen. Rel. Grav.}{#1}}
\def\JETF#1 {\Jl{Zh. Eksp. Teor. Fiz.}{#1}}
\def\JETP#1 {\Jl{Sov. Phys. JETP}{#1}}
\def\JHEP#1 {\Jl{JHEP}{#1}}
\def\JMP#1 {\Jl{J. Math. Phys.}{#1}}
\def\NPB#1 {\Jl{Nucl. Phys. B}{#1}}
\def\NP#1 {\Jl{Nucl. Phys.}{#1}}
\def\PLA#1 {\Jl{Phys. Lett. A}{#1}}
\def\PLB#1 {\Jl{Phys. Lett. B}{#1}}
\def\PRD#1 {\Jl{Phys. Rev. D}{#1}}
\def\PRL#1 {\Jl{Phys. Rev. Lett.}{#1}}

\def\lal{&&\nqq {}}

\def\beq{\begin{equation}}
\def\eeq{\end{equation}}
\def\bear{\begin{eqnarray}}
\def\bearr{\begin{eqnarray} \lal}
\def\ear{\end{eqnarray}}
\def\earn{\nonumber \end{eqnarray}}

\def\e{{\,\rm e}}

\documentclass[%
 reprint,
 amsmath,amssymb,aps,onecolumn
]{revtex4-2}

\usepackage{dcolumn}
\usepackage{bm}
\usepackage{wrapfig}
\usepackage[english]{babel}
\usepackage[utf8]{inputenc}
\usepackage{amsthm}
\usepackage{mathtools}
\usepackage{physics}
\usepackage{xcolor}
\usepackage{graphicx}
\usepackage[framemethod=tikz]{mdframed}
\usepackage[left=15mm,right=15mm,top=20mm]{geometry} 
\usepackage{adjustbox}
\usepackage{placeins}
\usepackage[T1]{fontenc}
\usepackage{lipsum}
\usepackage{csquotes}
\usepackage{tensor}
\usepackage{tikz}
\usepackage{tabularx}
\usepackage{float}
\usepackage[hyperindex,colorlinks]{hyperref}

\begin{document}


\title{Black hole and wormhole solutions in Einstein-Maxwell-Scalar theory}

\author{Júlio C. Fabris}
 \email{julio.fabris@cosmo-ufes.org}
\author{Tales A. O. Gomes}%
 \email{talesaugustog@gmail.com}
 
 \author{Denis C. Rodrigues}
 \email{deniscr@gmail.com}
 
\affiliation{%
 Núcleo Cosmo-Ufes {\&} Departamento de Física, Universidade Federal do Espírito Santo (UFES), Av. Fernando Ferrari, 540, Vitória 29075-910, Brazil.
}%

\date{\today}

\begin{abstract}
    
We classified and studied the charged black hole and wormhole solutions in the Einstein--Maxwell system in the presence of a massless, real scalar field. The possible existence of charged black holes in general scalar--tensor theories was studied in {Bronnikov et al, 1999;}
black holes and wormholes exist for a negative kinetic term for the scalar field. Using a conformal transformation, the static, spherically symmetric possible structures in the minimal coupled system are described. Besides wormholes and naked singularities, only a restricted class of black hole exists, exhibiting a horizon with an infinite surface and a timelike central singularity. The black holes and wormholes defined in the Einstein frame have some specificities with respect to the non-minimal coupling original frame, which are discussed in \mbox{the text.}
 
 \vspace{1em}
 \end{abstract}

 
\maketitle


\section{Introduction} \label{sec:introduction}

Black holes (BHs) are objects predicted by the general relativity (GR) theory. Their main characteristic feature is the existence of an event horizon, a~hypersurface separating two regions, the~internal one, which generally contains a singularity, and~the external one with the asymptotic spatial infinity where the observer may be located. Even if physically and conceptually a BH may be considered an intriguing object, with~many subtle properties, its characterization is very simple. It is conjectured that only three parameters are enough to describe a BH: its mass, charge, and angular momentum. This is the main content of the no hair conjecture: just three numbers define the BH. In~this sense, the BH may be considered as a simpler object existing in nature. However, the possibility of existing BHs with additional parameters, for~example a~scalar charge~\cite{herdeiro1,japa1}, has been evoked in the literature, but~the stability of the resulting structure is not always assured. Besides~BHs, GRs admit solutions with no horizon and no singularity and containing two asymptotic spatial regions connected by a throat in spacetime. They have been named wormholes (WHs) since they exhibit a short path between two distant regions. In~general, a WH requires repulsive effects, at~least around the throat. Repulsive gravitational effects may be a problem for its stability and also for the possibility to be traversable, especially by a human being. It must be stressed that the stability of a BH is also frequently problematic, especially when it contains some kind of hair. However, the~classical BH configurations represented by the Schwarzschild solution (which is static and uncharged), the Reissner--Nordstr\"om solution (static and charged), and the Kerr solution (rotating and uncharged) are stable, except~perhaps for their corresponding extreme cases. If~in some solutions, the~event horizon is absent, the~singularity becomes visible to an external observer. This configuration is called a naked singularity (NS). They are in general unstable, meaning that they may not exist in nature. The~conjecture asserting that a naked singularity may not exist in nature is dubbed cosmic censorship. For~some extensive descriptions of these objects and their properties, see~\cite{chandra,visser,frolov}.

BHs have been considered for a long period as hypothetical objects. This situation has changed drastically in recent times. The~detection by the LIGO project~\cite{ligo} of gravitational waves emitted by the coalescence of two BHs and the first image of a BH revealed by the Event Horizon Telescope (EHT) \cite{eht} have left little doubt about the existence of these objects. The~characteristics of the emissions coming from galaxies with an active nucleus (AGN) show also clear evidence of the existence of supermassive black holes in their center. The~possibility of the existence of a central supermassive BH has been extended essentially to almost all galaxies, including our galaxy. The~mass of an isolated BH (called an astrophysical BH, resulting from the collapse of massive stars) is of the order of some tens of solar masses ($M_\odot$), but~the mass of the BHs in the center of galaxies spans from millions to billions of solar mass. While the BHs have mounted to the status of a legitimate astrophysical object, WHs remain a mathematical result of the GR theory, with~no clear evidence until now for their~existence.
 
In the literature, there is a plethora of BH solutions found in many different contexts, most of them extending the content of known classical solutions. Many
different fields have been considered as a source for the BH structure. The~classical BH solutions are vacuum solutions for massive, charged or uncharged, rotating or non-rotating configurations. In~these classical solutions, only the RN solution requires a non-trivial right-hand side of the GR equations due to the presence of the electromagnetic field. Other possible energy--momentum tensors, containing gauge fields or scalar fields, may be introduced (see~\cite{herdeiro2} and references therein), leading generally to new BH or even WH structures. Furthermore, non-minimal coupling between the fields and/or with the geometry may be considered, leading to quite rich configurations. Theories with extra dimensions may lead also to quite special structures.
However, the~stability of each of these configurations must be tested carefully. See~\cite{Bronnikovbook} for a description of some of these~structures.

Scalar fields are the simpler source that can be added in a gravitational theory. However, scalar fields may favor the appearance of NSs instead of BHs, unless~this appears as a phantom field, that is a~field with negative kinetic energy, or~if it is coupled in a non-trivial way with gravity and/or other gauge fields.
An example is the Einstein--Maxwell dilaton (EMD) system: they contain BHs that are asymptotically flat, but only in the case that the scalar field is coupled with
the Maxwell field or if the scalar field is phantom~\cite{emd1,emd2,emd3,gmf}. 
Of course, the~presence of self-interaction for the scalar field, represented by a potential term, $V(\phi)$, may change all these considerations. However, it is not always clear how to motivate the presence of the potential term. In~what follows, we ignore this~possibility.

BHs and WHs in the Brans--Dicke theory in the presence of an EM field were extensively studied in~\cite{Bronnikov99}. In~this theory, the~EM field is coupled minimally to gravity, but~the scalar field is coupled non-minimally. All possible BH, WH, and NS cases were identified. The~BH solutions belong to the so-called cold BHs: the horizon has an infinite area, and its corresponding Hawking temperature is zero. They are in general unstable~\cite{bfz}, even if they present some interesting causal structures. Cold black hole are also present when the EM field is absent in the non-minimal and minimal coupling of the scalar field with gravity~\cite{Bronnikovstr,Bronnikovsta,Bronnikovcoldbh}. However, in~all cases, the~scalar field must have a phantom character when transposed to the Einstein~frame.

The goal of the present work is to revisit the problem treated in~\cite{Bronnikov99}. The~main difference is that the BH and WH solutions are identified in the Einstein frame. Again, BHs and WHs will be found only if the scalar field is phantom; otherwise, only NSs appear. To~our knowledge, the BH and WH solutions identified here are new, even if some of their features are similar to other known solutions, especially those found in~\cite{Bronnikov99}. Besides, to~complement the analysis presented in~\cite{Bronnikov99}, our goal was to stress the properties of the solutions with their physical content and the role played by the conformal transformation, and~its main consequences, in~passing from the Jordan to the Einstein frame.
For example, in~contrast to the original Brans--Dicke case, with~its different types of BH structures, there are only two possible classes of BHs, one with a double horizon and another one with a single horizon, characteristics similar to the Reissner--Nordstr\"om BH, non-extreme and extreme. In~both cases, the~central singularity is timelike (hence, repulsive), which may again be related to the presence of the Maxwell field.
Both solutions are cold black holes, which may be related to the presence of a (phantom) scalar field. The~WH class of solutions requires also a phantom scalar field, but~it is regular without horizons or~singularities. 

In the next section, we derive the general solution. All derivations are very similar to the calculations exhibit in~\cite{Bronnikov99}. In~Section~\ref{sec3}, the~different solutions 
are classified. In~Section~\ref{bhs}, the~NS, BH, and WH solutions are identified. In~Section~\ref{sec:conclusions}, we present our~conclusions.

\section{Field Equations and General~Solutions} \label{sec:fieldseq}

Our starting point is the action of the Einstein--Maxwell theory with a minimally coupled massless scalar field,
\begin{equation} \label{action}
 S = \int d^4x \sqrt{-g}(R - \epsilon\nabla_\alpha\Phi\nabla^{\alpha}\Phi + F_{\alpha\beta}F^{\alpha\beta}),
\end{equation}
where $R$ is the Ricci scalar, $\Phi$ is a massless, real scalar field, and $F^{\mu\nu}$ is Maxwell's tensor. Our conventions are the following: the metric signature is $(+ - - -)$; the Ricci tensor is $R_{\mu\nu} = \partial_\rho\Gamma^\rho_{\mu\nu} - \partial_\nu\Gamma^\rho_{\mu\rho} + \Gamma^\rho_{\mu\nu} \Gamma^\sigma_{\rho\sigma} - \Gamma^\rho_{\mu\sigma} \Gamma^\sigma_{\nu\rho}$. Using the canonical expression for the energy--momentum tensor,
\begin{eqnarray}
T_{\mu\nu} = - \frac{2}{\sqrt{-g}}\frac{\delta \sqrt{-g}{\cal L}}{\delta g^{\mu\nu}},
\end{eqnarray}
 we obtain the energy--momentum tensor for the scalar field and for the electromagnetic field, respectively:
\begin{eqnarray}
T^\Phi_{\mu\nu} &=& \epsilon \biggr(\Phi_{;\mu}\Phi_{;\nu} - \frac{1}{2}g_{\mu\nu}\Phi_{;\rho}\Phi_{;\rho}\biggl),\\
T^F_{\mu\nu} &=& - 2\biggr(F_{\mu\rho}F_\nu^\rho - \frac{1}{4}g_{\mu\nu}F_{\rho\sigma}F^{\rho\sigma}\biggl).
\end{eqnarray}

The parameter $\epsilon$ can assume the values $+1$, which gives us a positive kinetic energy for the scalar field, which is called canonical, and~$-1$, which represents the ``phantom'' scalar field with negative kinetic energy. This action describes the interaction between gravity and a massless scalar field in the presence of an electromagnetic field in four dimensions. The~field equations generated by this action are:
\begin{equation}
 R_{\mu\nu} = \epsilon\nabla_{\mu}\Phi\nabla_{\nu}\Phi - \left(2F_{\mu\lambda}F^{\lambda}_{\nu} - \frac{1}{2}g_{\mu\nu}F_{\alpha\beta}F^{\alpha\beta}\right),
\end{equation}
\begin{equation}
\label{eq:maxwell}
 \nabla_{\alpha}F^{\alpha\beta}=0,
\end{equation}
\begin{equation}
 \nabla_{\alpha}\left[\frac{1}{2}\epsilon^{\alpha\beta\lambda\sigma}F_{\lambda\sigma}\right]=0,
\end{equation}
\begin{equation}
\label{eq:kleingordon}
 \nabla_{\alpha}\nabla^{\alpha}\Phi=0,
\end{equation}
where $\epsilon^{\alpha\beta\lambda\sigma}$ is the Levi-Civita~symbol. 

Before going on to the specific computations, we provide some words on the choice of the action (\ref{action}). Our goal is to explore the black hole and wormhole structures in minimally coupled scalar and electromagnetic fields. In~doing so, we neglected two interesting cases. The~first is the
addition of a self-interaction in the scalar sector that can be represented by a potential term $V(\Phi)$. The~reason was two-fold. First, we intended to keep contact with the study carried out in~\cite{Bronnikov99}, where a similar analysis was
carried out in the Jordan frame, in~the context of the Brans--Dicke theory, which in its traditional form has no potential for the scalar field. The~Brans--Dicke structure is connected to our action (\ref{action}) by a conformal transformation. Moreover, the~presence of a potential, even in its simpler form representing a massive field, spoils the possibility to have analytical solutions, and~a quite sophisticated numerical treatment is necessary, an~approach that lies beyond the scope of the present work. However, we stress the important possible connections that the case with a potential $V(\Phi)$ has, for~example, with~the
$f(R)$ theories, which can be reformulated in a Brans--Dicke-type theory with a suitable potential~\cite{f(R)}. Another possibility, in~introducing a self-interaction in the scalar sector, is the inverse problem: the potential is determined from a given solution, with~some specific features. This problem is interesting also, but~the motivation for the potential determined in this way is not always~clear.

The second important restriction of the action (\ref{action}) is the absence of a coupling between the scalar field and the electromagnetic field. This case was extensively studied in~\cite{emd1,emd2,emd3,gmf}. These studies have explored all possible solutions (always in the absence of a potential term), asymptotically flat and non-flat, in~the Einstein frame. In~this sense, perhaps it would be interesting to revert the problem and write the solutions in the original non-minimal coupling (sometimes also called the string frame). However, this would constitute another different problem, deserving a separate~analysis.

Coming back to the action (\ref{action}), in~deriving the corresponding static, spherically symmetric solutions, we followed closely the computation carried out in~\cite{Bronnikov99}. 
We give some details of this computation in order to be as complete as~possible.

We considered a static, spherically symmetric spacetime described by the metric,
\begin{equation}
\label{eq_01}
 ds^2=\e^{2\gamma}dt^2-\e^{2\alpha}du^2-\e^{2\beta}d\Omega^2,
\end{equation}
where $\gamma$, $\alpha$, and $\beta$ are functions of the radial coordinate $u$ only and $d\Omega^2=d\theta^2+\sin^2\theta d\phi^2$ is the differential two-sphere. The~non-vanishing terms of Ricci tensor for this metric are,
\begin{align}
 R_{00}&=\e^{-2\alpha+2\gamma}[\gamma''+(\gamma'-\alpha'+2\beta')\gamma'],\\
 R_{11}&= -\gamma''-2\beta''+(2\beta'+\gamma')\alpha'-2(\beta')^2-(\gamma')^2,\\
 R_{22}&= 1+\e^{-2\alpha+2\beta}[(\alpha'-\gamma'-2\beta')\beta'-\beta''],
\end{align}
where primes denote the derivative with respect to the radial coordinate $u$. 

For a point charge in a static spherically symmetric spacetime, the only non-null terms of Maxwell's tensor are,
\begin{equation}
 F_{10} = -F_{01}= E(u).
\end{equation}

Therefore,~Maxwell's Equation~(\ref{eq:maxwell}) gives us,
\begin{equation}
\label{electricfield}
 E(u)=Q \e^{\alpha+\gamma-2\beta},
\end{equation}
where $Q$ is a constant of integration and~can be interpreted as the electric charge. On~the other hand, since we are working with a static, spherically symmetric spacetime, we can assume that the scalar field is a function of the radial coordinate only ($\Phi\equiv\Phi(u)$). Thus, we obtain,
\begin{equation}
\label{eq:scalar'}
 \Phi'=C\e^{\alpha-\gamma-2\beta},
\end{equation}
where $C$ is a constant of integration and~may be interpreted as a scalar~charge. 

The resulting equations, using the non-vanishing Ricci tensor components and the solution for the electric field, are:
\begin{align}
 \label{eq:13}\gamma''+(\gamma'-\alpha'+2\beta')\gamma'&=Q^2\e^{2\alpha-4\beta},\\[2ex]
 \gamma''+2\beta''-(2\beta'+\gamma')\alpha'+2\beta'^2+\gamma'^2&=-\epsilon C^2\e^{2\alpha-2\gamma-4\beta}+Q^2\e^{2\gamma-4\beta},\\[2ex]
 \beta''-(\alpha'-\gamma'-2\beta')\beta'&=\e^{2\alpha-2\beta}-Q^2\e^{2\gamma-4\beta}.
\end{align}

Equation~(\ref{eq:13}) suggests that a suitable choice of the coordinates is the harmonic coordinates, where $\alpha(u)=\gamma(u)+2\beta(u)$. The~harmonic coordinates are generally a convenient choice when scalar fields are present. With~this choice, the~equations become:
\begin{equation}
\label{r00eq}
 \gamma''=Q^2\e^{2\gamma},
\end{equation}
\begin{equation}
\label{r11eq}
 \beta''-\beta'^2-2\beta'\gamma'=-\frac{1}{2}\epsilon C^2,
\end{equation}
\begin{equation}
\label{r22eq}
 \gamma''+\beta''=\e^{2\gamma+2\beta}.
\end{equation}

This last equation can be integrated, leading to,
\begin{equation}
 (\gamma'+\beta')^2=\e^{2\gamma+2\beta}+k^2{\rm sign}\, k,
\end{equation}
where $k$ is a constant of integration. Making the substitution $\gamma+\beta=-\ln[s(u)]$, we have:
\begin{equation}
 s'^2=1+k^2{\rm sign}\, k \, s^2.
\end{equation}

The solution for $s(u)$ will depend on the sign of $k$; thus, we may write $s(u)\equiv s(k,u)$, and the solution will be:
\begin{equation}
\label{seq}
s(k,u)=\begin{cases} k^{-1}\sinh(ku), &k>0,\\
u, &k=0,\\
k^{-1}\sin(ku), &k<0.
\end{cases}
\end{equation}

The solution for $\gamma$ is obtained by integrating Equation~(\ref{r00eq}), which leads to,
\begin{equation}
 \gamma'^2=Q^2\e^{2\gamma} + \lambda^2{\rm sign}\, \lambda,
\end{equation}
where $\lambda$ is another constant of integration. To~solve this equation, we use the substitution $\gamma(u)=-\ln[h(u)]$. Thus,
\begin{equation}
\label{gaso}
 h'^2=Q^2+\lambda^2{\rm sign}\, \lambda\, h^2.
\end{equation}

Comparing this equation to {Equation}
~(\ref{seq}), we obtain the relation $h(u)=Qs(\lambda,u+u_0)$. 

Finally, the~general solution for the metric will be,
\begin{equation}
 \label{gmetric}
 ds^2=\frac{Q^{-2}dt^2}{s^2(\lambda,u+u_0)}-Q^2\frac{s^2(\lambda,u+u_0)}{s^2(k,u)}\left(\frac{du^2}{s^2(k,u)}+d\Omega^2\right).
\end{equation}

The definition of the function $s(\lambda, u + u_0)$ is similar to the definition of $s(k,u)$ given~above.

From Equation~(\ref{r11eq}), we have the following relation for the integration constant:
\begin{equation}
\label{sire}
 k^2 {\rm sign}\, k - \lambda^2 {\rm sign}\,\lambda=\frac{\epsilon C^2}{2}.
\end{equation}

The scalar and electromagnetic fields are written, respectively, as:~\begin{equation}
\label{eq:fieldssol}
 \Phi(u)=Cu, \quad \mbox{and}\quad E(u)=\frac{1}{Qs^2(\lambda,u+u_0)}.
\end{equation}

Without losing generality, we can normalize $g_{00}=1$ at $u = 0$ by imposing \mbox{the condition,}
\begin{equation}
\label{con}
 s^2(\lambda,u_0)=\frac{1}{Q^2}.
\end{equation}

We thus have four integration constants: $\lambda$, $k$, $Q$, and $C$. Another important constant is the geometric mass $m$ of the configuration, which we can obtain by comparing the asymptotic {Equation}~(\ref{gmetric}) with the Schwarzschild metric,
\begin{equation}
 \gamma_{sc}(r)=\frac{1}{2}\ln(1-\frac{2m}{r}).
\end{equation}

At the asymptotic $u\rightarrow 0$, it behaves as $r^{-1}$. Hence, 
\begin{equation*}
 \gamma_{sc}(u)=\frac{1}{2}\ln(1-2mu)\quad\rightarrow\quad\gamma_{sc}'(0)=-m,
\end{equation*}
and comparing with {Equation}~(\ref{gmetric}),
\begin{equation}
\label{mre}
 \gamma'(u)=-\frac{s'(\lambda,u+u_0)}{s(\lambda,u+u_0)}\Rightarrow m =\frac{s'(\lambda,u_0)}{s(\lambda,u_0)}.
\end{equation}

Now, we can connect all constants using the asymptotic behavior of (\ref{gaso}) as $u\rightarrow 0$, the~relations (\ref{sire},\ref{mre}), and the condition (\ref{con}), obtaining,
\begin{equation}
\label{pre}
 m^2-Q^2=\lambda^2{\rm sign}\,\lambda= k^2{\rm sign}\,k -\frac{\epsilon C^2}{2}.
\end{equation}

Finally, we can write the metric in terms of $s(k,u)$ only, as:
\begin{equation}
\label{fm}
 ds^2=\frac{s^2(\lambda,u_0)dt^2}{s^2(\lambda,u+u_0)}-\frac{s^2(\lambda,u+u_0)}{s^2(\lambda,u_0)s^2(k,u)}\left(\frac{du^2}{s^2(k,u)}+d\Omega^2\right).
\end{equation}

From here, the~Reissner--Nordstr\"om solution of GR is recovered by putting $C=0$; thus, $\lambda=k$. In~this sense, we have three possible cases, $\lambda=k>0$, $\lambda=k=0$, and $\lambda=k<0$, corresponding to the RN non-extreme ($m>Q$), extreme ($m=Q$), and naked singularity ($m<Q$), respectively. In~each case, the~familiar form of the RN solution is obtained with the transformation:
\begin{equation}
 r=\frac{s(\lambda,u+u_0)}{s(\lambda,u_0)s(k,u)}.
\end{equation}

Another interesting limiting case is for $Q=0$, which is the scalar--vacuum solution, obtained in~\cite{bronnikov73}. The~limit $Q\rightarrow 0$ must be taken preserving the boundary condition (\ref{con}). This leads to:
\begin{equation}
 \lambda\geq0\quad\mbox{and}\quad u_0\rightarrow\infty\quad\Rightarrow\quad s(\lambda,u+u_0)\rightarrow e^{2\lambda u}.
\end{equation}

This solution was studied in detail in~\cite{Bronnikovstr,Bronnikovsta}.

Noticing that for $u\rightarrow 0$, we have $s(k,0)\rightarrow 0$ for all $k$, we can identify two surfaces of interest from the metric (\ref{fm}). One surface is at $u=-u_0$, where the metric terms behave as:
\begin{equation*}
 \e^{2\gamma}\rightarrow\infty,\quad \mbox{and} \quad \e^{2\alpha}, \e^{2\beta}\rightarrow 0,
\end{equation*}
characterizing a central singularity, and the other is At $u=0$, where:
\begin{equation*}
 \e^{2\gamma}\rightarrow 1,\quad \mbox{and} \quad \e^{2\alpha}, \e^{2\beta}\rightarrow \infty ,
\end{equation*}
that is an asymptotic flat surface (Minkowski) expressed in the harmonic coordinates. Therefore, the~general solution (\ref{fm}) will have at least one singularity and one asymptotic flat surface. The~sign of $\lambda$ and $k$ will determine the combination of functions in the metric, according to {Equation}~(\ref{seq}). The~structure of the spacetime may also have one or more horizons, or~even no horizon at all. In~all possible cases, the~solutions that we look for are BHs and~WHs.
 
 \section{Classifying the Static, Spherically Symmetric~Solutions}\label{sec3}

The metric (\ref{fm}) has different forms depending on the sign of the constant of integration $\lambda$ and $k$. Each combination of signs will provide a relation between the constants. In~this section, we investigate each case representing an independent solution. All possible relations are listed in Tables~\ref{table:1} and \ref{table:2}. The~features of these different cases depend on, crucially, whether~the scalar field is canonical ($\epsilon = 1$, Table~\ref{table:1}) or
if it is phantom ($\epsilon = - 1$, Table~\ref{table:2}).

\begin{table}[H]
\caption{Possible relations between the constants of integration in the canonical sector, $\epsilon=+1$. The~$*$ indicates the cases that give complex values for the constants, which are not of~{interest.} 
}
\label{table:1}
\centering
\begin{tabular}{| m{2.77cm}|| m{3cm}|m{3.2cm}|m{3.2cm}|} 
\hline
$\epsilon=+$& $k=0$ & $k>0$ & $k<0$\\ 
\hline\hline
$\lambda=0$ & $C=0$ & $k^2=\frac{C^2}{2}$ & * $k^2=-\frac{C^2}{2}$ \\ 
\hline
$\lambda>0$ & * $\lambda^2=-\frac{C^2}{2}$ & $k^2-\lambda^2=\frac{C^2}{2}$ & * $k^2+\lambda^2=-\frac{C^2}{2}$ \\
\hline
$\lambda<0$ & $\lambda^2=\frac{C^2}{2}$ & $k^2+\lambda^2=\frac{C^2}{2}$ & $\lambda^2-k^2=\frac{C^2}{2}$ \\
\hline
\end{tabular}
\end{table}
\unskip

\begin{table}[H]
\centering
\caption{Possible relations between the constants for the phantom sector, $\epsilon=-1$. $*$ indicates the cases that give complex values for the constants, which are not of~{interest.}}
\label{table:2}
\begin{tabular}{| m{2.77cm}|| m{3cm}|m{3.2cm}|m{3.2cm}|} 
\hline
$\epsilon=-$ & $k=0$ & $k>0$ & $k<0$\\ 
\hline\hline
$\lambda=0$ & $C=0$ & * $k^2=-\frac{C^2}{2}$ & $k^2=\frac{C^2}{2}$ \\ 
\hline
$\lambda>0$ & $\lambda^2=\frac{C^2}{2}$ & $\lambda^2-k^2=\frac{C^2}{2}$ & $k^2+\lambda^2=\frac{C^2}{2}$ \\
\hline
$\lambda<0$ & * $\lambda^2=-\frac{C^2}{2}$ & * $k^2+\lambda^2=-\frac{C^2}{2}$ & $k^2-\lambda^2=\frac{C^2}{2}$ \\
\hline
\end{tabular}
\end{table}

Assuming all constants to be real, some of these relations will not be valid since they provide complex values for some of the constants. Excluding all the cases where the constants assume complex values, we are left with the independent solutions, which are listed in Tables~\ref{table:3} ($\epsilon = 1$) and \ref{table:4} ($\epsilon = - 1$).

\unskip

\begin{table}[H]
\caption{Possible independent solutions for $\epsilon=+1$.}
\label{table:3}
\centering
\setlength{\tabcolsep}{20mm}\begin{tabular}{|c |} 
 \hline
\boldmath{ $\epsilon=+1$ }\\ 
 \hline
 $\lambda>0$ and $k>0$ \\
 \hline
 $\lambda<0$ and $k>0$ \\
 \hline
 $\lambda<0$ and $k=0$\\
 \hline
 $\lambda<0$ and $k<0$ \\
 \hline
 $\lambda=0$ and $k>0$ \\ 
 \hline
\end{tabular}
\end{table}
\unskip

\begin{table}[H]
\caption{Possible independent solutions for $\epsilon= - 1$.}
\label{table:4}
\centering
\setlength{\tabcolsep}{20mm}\begin{tabular}{|c |} 
 \toprule
 \boldmath{$\epsilon=-1$} \\ 
 \hline
 $\lambda>0$ and $k>0$ \\
 \hline
 $\lambda>0$ and $k=0$ \\
 \hline
 $\lambda>0$ and $k<0$\\
 \hline
 $\lambda<0$ and $k<0$ \\
 \hline
 $\lambda=0$ and $k<0$ \\ 
 \hline
\end{tabular}
\end{table}
\unskip

\subsection{Independent Solutions in the Canonical~Sector}
\label{seq:canonical}

The canonical sector is given by a positive value for the kinetic energy of the scalar field, $\epsilon=+1$. In~this section, we analyze each independent solution of {Equation}~(\ref{fm}) listed in Table~\ref{table:3}, knowing that all solutions have the singular surface at $u=-u_0$ and are asymptotically flat at $u=0$. Thus, we~have:
\begin{enumerate}
 \item $k>\lambda>0$: In this case, the~metric takes the form,
\begin{equation}
 \label{m+1}
  ds^2=\frac{\lambda^2\sinh^2(\lambda u_0)dt^2}{\sinh^2[\lambda(u+u_0)]}-\frac{k^2\sinh^2[\lambda(u+u_0)]}{\lambda^2\sinh^2(\lambda u_0)\sinh^2(ku)}\left(\frac{k^2du^2}{\sinh^2(ku)}+d\Omega^2\right),
 \end{equation}
 and the constants are related by:
\begin{equation}
 \label{rel+1A}
  m^2-Q^2=\lambda^2= k^2-\frac{C^2}{2}.
 \end{equation}
 
 Therefore, we can see that for $u\rightarrow\infty$, we have $\e^{\gamma}\rightarrow0$. 
 Thus, there may be a horizon at this surface. However,~analyzing the angular term of the metric, for~large values of $u$, it can be approximated to:
 \begin{equation*}
  \e^{2\beta}\approx \e^{2(\lambda-k)u},
 \end{equation*}
 and since $k>\lambda$, we have,
\begin{equation}
  \lim_{u\rightarrow\infty}\e^{2\beta(u)}=0.
 \end{equation}
 
 Hence, we have another singular surface. This case can only describe a naked singularity with an asymptotic flat~spacetime;

 \item $k>0>\lambda$: Using the appropriate functions for this case, the metric is written as,
\begin{equation}
  ds^2=\frac{\sin^2(\lambda u_0)dt^2}{\sin^2[\lambda(u+u_0)]} -\frac{k^2\sin^2[\lambda(u+u_0)]}{\sin^2(\lambda u_0)\sinh^2(ku)}\left(\frac{k^2du^2}{\sinh^2(ku)}+d\Omega^2\right),
 \end{equation}
 where:
\begin{equation}
  m^2-Q^2=-\lambda^2= k^2 -\frac{C^2}{2}.
 \end{equation}
 
 Because the sine function has a finite range and~it is a periodic function, there is no surface where $\e^{\gamma}\rightarrow 0$; thus, it has no horizon, but~it has many singular points. The~singular points of the first term of the metric are at,
\begin{equation}
  u_{\lambda,n}=\pm\frac{\pi n}{\lambda}-u_0 ,\quad n=0,1,2,3...
 \end{equation}
 
 Other interesting surfaces are at $u\rightarrow\pm\infty$, where we have: \begin{equation}
  \lim_{u\rightarrow\pm\infty}\e^{2\beta(u)}=0,
 \end{equation}
 which describes another singular surface. This case has only naked~singularities;

 \item $\lambda<k=0$: The metric in this case takes the form,
\begin{equation}
  ds^2=\frac{\sin^2(\lambda u_0)dt^2}{\sin^2[\lambda(u+u_0)]} -\frac{\sin^2[\lambda(u+u_0)]}{\sin^2(\lambda u_0)u^2}\left(\frac{du^2}{u^2}+d\Omega^2\right).
 \end{equation}
 
 The constants of integration are related by,
\begin{equation}
  Q^2-m^2=\lambda^2=\frac{C^2}{2}.
 \end{equation}
 
 This case is exactly the same as the previous one, where there is no horizon, and~we also have the same singular points and local~minimums; 
  
 \item $\lambda<k<0$: Writing the metric for this case, we find,
\begin{equation}
  ds^2=\frac{\sin^2(\lambda u_0)dt^2}{\sin^2[\lambda(u+u_0)]}-\frac{k^2\sin^2[\lambda(u+u_0)]}{\sin^2(\lambda u_0)\sin^2(ku)}\left(\frac{k^2du^2}{\sin^2(ku)}+d\Omega^2\right),
 \end{equation}
 with the relation,
\begin{equation}
  Q^2-m^2=\lambda^2=k^2+\frac{C^2}{2}.
 \end{equation}

 Again, the~first term of the metric $\e^{2\gamma}$ is the same as the last cases, but~the angular term now has another sine function. Therefore, we still have the singular points $u_{s,n}$, but~now, we have different local minimums. Another interesting surface in this case~is,
\begin{equation}
  u_{k,n}=\pm\frac{\pi n}{k}, \quad n=0,1,2,3...
 \end{equation}
 where,
\begin{equation}
  \lim_{u\rightarrow u_{k,n}}\e^{2\beta}\rightarrow\infty,
 \end{equation}
 indicating an asymptotic surface. Since, $|k|<|\lambda|$, we have that the interval between the singular points $u_{s,n}$ is smaller than the interval between the asymptotic regions $u_{k,n}$. Hence,
\begin{equation}
  |u_{\lambda,n+1}-u_{\lambda,n}|<|u_{k,n+1}-u_{k,n}|,
 \end{equation}
 and there is always a singular surface between the asymptotic surfaces. Thus, this can only be a naked~singularity.

 \item $k>\lambda=0$: Here, the metric is written as,
\begin{equation}
  ds^2=\frac{u_0^2dt^2}{(u+u_0)^2}
  -\frac{k^2(u+u_0)^2}{u_0^2\sinh^2(ku)}\left(\frac{k^2du^2}{\sinh^2(ku)}+d\Omega^2\right),
 \end{equation}
 with the relation,
\begin{equation}
  m^2-Q^2= k^2-\frac{C^2}{2}=0.
 \end{equation}
 Here, we can identify that $\e^{\gamma}\rightarrow 0$ in the surface $u=\pm\infty$, but~one can verify that at this surface, the third criterion for BH selection is not satisfied. Moreover, $\e^{2\beta} \rightarrow 0$ for those points. Thus, this is not a horizon. This solution describes also a naked~singularity.
  
\end{enumerate}

\subsection{Independent Solutions in the Phantom~Sector}
\label{seq:phantom}

In this section, we analyze the phantom sector, where the kinetic energy of the scalar field has negative energy, using the possible solutions listed in Table~\ref{table:4}: 
\begin{enumerate}
 \item $\lambda>k>0$: In this case, we have the metric as,
\begin{equation}
\label{s-1}
 ds^2=\frac{\sinh^2(\lambda u_0)dt^2}{\sinh^2[\lambda(u+u_0)]}-\frac{k^2\sinh^2[\lambda(u+u_0)]}{\sinh^2(\lambda u_0)\sinh^2(ku)}\left(\frac{k^2du^2}{\sinh^2(ku)}+d\Omega^2\right),
\end{equation}
with the relation:
\begin{equation}
\label{rel7.1}
 m^2-Q^2=\lambda^2= k^2+\frac{C^2}{2}.
\end{equation}
We can see that for $u\rightarrow \pm\infty$, we have $\e^{\gamma}\rightarrow 0$, which is one condition for having a horizon. Analyzing the angular term for large values of $u$, we can approximate it as:
\begin{equation*}
 \e^{2\beta}\approx \e^{2(\lambda-k)u}.
\end{equation*}
Since we have that $\lambda>k$, the~surface $u\rightarrow-\infty$ implies $\e^{2\beta}\rightarrow 0$. Therefore, it is a singular surface rather than a horizon. On~the other hand, for~$u\rightarrow\infty$, we have $\e^{\beta}\rightarrow \infty$, which characterizes a horizon with an infinite surface area. Then, working in the range $-u_0<u<\infty$, this solution indeed represents a BH. This solution can be mapped to one of the charged Brans--Dicke BH (Case [1-] of~\cite{Bronnikov99}) through a conformal transformation. This is expected since the conformal transformation that maps the Einstein frame in the Jordan frame is given by an exponential of the scalar field $\Phi$, being regular. However, the~overall structure and features of the solution are affected by this conformal~transformation.

In fact, the~conformal transformation connects the metric in the Einstein frame (used here) and the metric in the Jordan frame, used in~\cite{Bronnikov99}, according to the relation,
\begin{eqnarray}
ds^2_J = \varphi^{-1}ds^2_E,
\end{eqnarray}
where the subscripts $E$ and $J$ indicate the Einstein and Jordan frames, respectively, and~$\varphi$ is the original Brans--Dicke field, which is related to the gravitational coupling. The~scalar field used here and the original Brans--Dicke field obey the \mbox{following relation:}
\begin{eqnarray}
\Phi = \sqrt{|\frac{3}{2} + \omega|}\ln\varphi.
\end{eqnarray}
In this way, the~metric in the original frame that corresponds to the metric (\ref{s-1}) in the Einstein frame takes the form,
\begin{eqnarray}
ds_J^2 = e^{- \frac{Cu}{\sqrt{|\frac{3}{2} + \omega|}}}\biggr\{\frac{\sinh^2(\lambda u_0)dt^2}{\sinh^2[\lambda(u+u_0)]}-\frac{k^2\sinh^2[\lambda(u+u_0)]}{\sinh^2(\lambda u_0)\sinh^2(ku)}\left(\frac{k^2du^2}{\sinh^2(ku)}+d\Omega^2\right)\biggl\}.
\end{eqnarray}
The overall conformal factor introduces, with~respect to the metric in the 
Einstein frame, the~constant $C$, implying new conditions to have a black hole, that is to~obtain $g_{00} \rightarrow 0$ with $g_{22} \neq 0$. Moreover, the~conditions to have an analytical extension of the metric beyond the horizon are affected by the presence of the parameter $C$, leading to a large spectrum of possibilities, with~different structures. This remark applies to other black hole and wormhole solutions discussed~above;
\item $\lambda>k=0$: The metric in this case is written as,
\begin{equation}
\label{s-2}
 ds^2=\frac{\sinh^2(\lambda u_0)dt^2}{\sinh^2[\lambda(u+u_0)]}-\frac{\sinh^2[\lambda(u+u_0)]}{\sinh^2(\lambda u_0)u^2}\left(\frac{du^2}{u^2}+d\Omega^2\right),
\end{equation}
where we have the relation:
\begin{equation}
 m^2-Q^2=\lambda^2=\frac{C^2}{2}.
\end{equation}
In the limit $u\rightarrow\pm\infty$, this case has $\e^{2\gamma}\rightarrow 0$ and $\e^{2\beta}\rightarrow \infty$. Therefore, for~the appropriate coordinate range, this solution represents a BH and~is connected, by~a conformal transformation, to~the other charged Brans--Dicke BH in the Jordan frame (Case [2-] of~\cite{Bronnikov99}). Again, the~conformal transformation maps the solution in one frame into the solution in the other~frame;
  
\item $\lambda>0>k$: For this case, the metric takes the form,
\begin{equation}
\label{solwhorm1}
 ds^2=\frac{\sinh^2(\lambda u_0)dt^2}{\sinh^2[\lambda(u+u_0)]}-\frac{k^2\sinh^2[\lambda(u+u_0)]}{\sinh^2(\lambda u_0)\sin^2(ku)}\left(\frac{k^2du^2}{\sin^2(ku)}+d\Omega^2\right),
\end{equation}
with:
\begin{equation}
\label{eq:relwh1}
 m^2-Q^2=\lambda^2=-k^2+\frac{C^2}{2}.
\end{equation}
The radial function, $\e^{2\alpha}$, connects two spatial infinities, which is a characteristic of WHs. Thus, for~an appropriate range, excluding the singularity at $u=-u_0$, this can be a~wormhole;

\item $k<\lambda<0$:

For this case we have,
\begin{equation}
\label{solwhorm2}
 ds^2=\frac{\sin^2(\lambda u_0)dt^2}{\sin^2[\lambda(u+u_0)]}-\frac{k^2\sin^2[\lambda(u+u_0)]}{\sin^2(\lambda u_0)\sin^2(ku)}\left(\frac{k^2du^2}{\sin^2(ku)}+d\Omega^2\right),
\end{equation}
with,
\begin{equation}
 Q^2-m^2=\lambda^2=k^2-\frac{C^2}{2}.
\end{equation}
The radial term of the metric function $\e^{2\beta}$ diverges at $u=\frac{n\pi}{k}$, with~$n$ any integer, while the first term, $\e^{2\gamma}$, is singular at $u=\frac{n\pi}{\lambda}-u_0$. Since $|k|>|\lambda|$, $\e^{2\beta}$ oscillates faster than $\e^{2\gamma}$, thus we can always choose a coordinate range where we have two spatial infinities and avoid the singularity at $u=\frac{n\pi}{\lambda}-u_0$, therefore a~WH~solution; 

\item $k<\lambda=0$: The last independent solution is,
\begin{equation}
\label{solwhorm3}
 ds^2=\frac{u_0^2dt^2}{(u+u_0)^2}
 -\frac{k^2(u+u_0)^2}{u_0^2\sin^2(ku)}\left(\frac{k^2du^2}{\sin^2(ku)}+d\Omega^2\right),
\end{equation}
with,
\begin{equation}
 Q^2-m^2=k^2 -\frac{C^2}{2}=0.
\end{equation}
As in the previous cases where we have a sine function in $\e^{2\beta}$, there are Minkowski asymptotics at $u=\frac{n\pi}{k}$ and horizons only at $u=\pm\infty$. Thus, in~the appropriate coordinate range, excluding the singularity at $u=-u_0$, this is a WH~solution.
  
\end{enumerate}

\section{\label{bhs}Black Hole and Wormhole~Solutions}

In this section, we discuss the solutions that indeed have a black hole and wormhole structure. The~BH solutions have event horizons with an infinite area, i.e.,~$g_{22}\rightarrow \infty$ as $u\rightarrow u_h$, which corresponds to a zero Hawking temperature, and~a finite proper time for infalling particles to attain the horizon (classified as type B1 BHs in~\cite{Bronnikov99}, where the type B2 BHs were also defined, for~which the horizon is at an infinite geodesic distance). In~general, the~form of these solutions is not familiar because these solutions are found in the harmonic gauge, while the form of the most well-known solutions are in the quasi-global gauge. To~study the geometry, it is helpful to introduce a new radial coordinate, which is in the quasi-global coordinate. Before~that, we revise the conditions required for a given solution to represent a BH or a~WH.

\subsection{Criteria for Black Hole and Wormhole~Selection}

\label{sec:crit}
BH solutions are spacetimes with a singular surface bounded by an event horizon, $u_h$. For~a static, spherically symmetric spacetime, as {Equation}~(\ref{eq_01}), we have the following criteria to select the event horizon; see the discussion in~\cite{Bronnikov99}. At~this surface, $u=u_h$, we~have:
\begin{enumerate}
  \item The timelike Killing vector becomes null, which means $\e^{\gamma}\rightarrow 0$;
  
  \item The surface area of the horizon is always positive non-null, so $\e^{2\beta}>0$;
  
  \item For an observer at rest, the~horizon is invisible; thus, the integral
  \begin{equation*}
   t^*=\int \e^{\alpha-\gamma}du\rightarrow \infty \quad\mbox{as}\quad u\rightarrow u_h;
  \end{equation*}
  
  \item The Hawking temperature $T_H$ is finite. One may use the following expression for the Hawking temperature of a surface $u=u_h$, in~natural units,
\begin{equation} 
  T_H=\lim_{u\rightarrow u_h}\frac{\kappa(u)}{2\pi}, \qquad \kappa(u)\overset{\rm def}{=}
  \e^{\gamma-\alpha}|\gamma'|;
 \end{equation}
  
  \item {The Kretschmann scalar $\mathcal{K}$ is finite. We discuss more about this scalar in Appendix~\ref{sec:appendix};}
  
  \item The metric must admit an analytical extension beyond the horizon.
 \end{enumerate}
 
 Wormholes are a spacetime containing two Minkowskian asymptotic regions connected by a throat, which is characterized by a minimum of the areal function
 $\e^{2\beta}$. The~spacetime must be geodesically complete and regular. This means that the Kretschmann scalar does not diverge anywhere. Moreover, this spacetime does not contain horizons, implying that it can be traversed from one asymptotic to the other, and~vice versa.

\subsection{First Black Hole~Solution}
\label{seq:first}
In Section~\ref{seq:phantom}, we verified that there are two possible BH solutions. The~first black hole solution is {Equation}~(\ref{s-1}). The~appropriate transformation for this solution is given by,
\begin{equation}
\label{eq:69}
 \e^{-2ku}=1-\frac{2k}{\rho}=P(\rho),
\end{equation}
which allows us to write the metric in a more familiar form,
\begin{equation}
\label{sol1}
 ds^2=f(\rho)dt^2-\frac{d\rho^2}{f(\rho)}-\frac{P(\rho)}{f(\rho)}\rho^2d\Omega^2,
\end{equation}
where:
\begin{equation}
\label{eq:conts}
 f(\rho)=\frac{4\lambda^2{P(\rho)}^a}{[(m+\lambda)-(m-\lambda){P(\rho)}^a]^2},\quad a=\frac{\lambda}{k}\quad\mbox{and}\quad m^2-Q^2=\lambda^2=k^2+\frac{C^2}{2}. 
\end{equation}

The geometric mass is defined as $m=\lambda\coth(\lambda u_0)$, and~since it must be always positive, we have that $u_0>0$. {The} 
 geometric mass defined previously coincides, following the usual definitions~\cite{wald}, with~the ADM mass, as can be explicitly verified. The~coordinate transformation used here changes the exponential character of the metric functions to a power law one. This transformation allows describing the interior region of the black hole. However,~this implies fractional powers of negative numbers, which violates the analyticity of the solution as the horizon is crossed. In~order to avoid the lost of analyticity, the allowed values for $a$ are:
\begin{equation}
\label{quantcond}
 a=1,2,3,4,5, . . . 
\end{equation}

The case $a=1$ corresponds to the non-extreme RN solution, since it leads to $C=0$. This ``quantization condition'' leads to a
discrete parametrization of the BH solutions and is also present, similarly, in~the Jordan frame~\cite{Bronnikov99}.
Using the solutions for the scalar and electric field {Equation}~(\ref{eq:fieldssol}) in the harmonic gauge and the coordinate transformation, {Equation}~(\ref{eq:69}), we have:
\begin{equation}
 \Phi(\rho)=-\frac{C}{2k}\ln|P(\rho)| \quad \mbox{and}\quad E(\rho)=Qf(\rho).
\end{equation}

Now, with~the metric in the quasi-global coordinate, one can directly identify one horizon at $\rho=2k$ and an asymptotic region at $\rho\rightarrow\infty$. We can also verify that $f(\rho)\rightarrow0$ as $\rho\rightarrow0$, giving us another~horizon.

The singularities in this coordinate system are given by the points where $f(\rho_s)\rightarrow\pm\infty$,
\begin{equation}
\label{eq:sing}
 {P(\rho_s)}^a=\left(1-\frac{2k}{\rho_s}\right)^a=\left(\frac{m+\lambda}{m-\lambda}\right).
\end{equation}

The right hand-side of the above equation is always positive, since by definition, $m>\lambda$. However, $P(\rho_s)$ can admit negative and positive roots or only positive roots depending on the value of $a$. Hence, depending on $a$, if~it is even or odd, the~singularity will have different positions, splitting the possible solutions into two cases. 
For both cases, the~singularity $f(\rho_s)\rightarrow+\infty$ is timelike, which is frequently a characteristic of charged~BHs.

\subsubsection{Case  of $a$ Odd}
\label{seq:aodd}
\begin{figure}[H]
\begin{minipage}{0.5\textwidth}
\centering
\includegraphics[width=8cm]{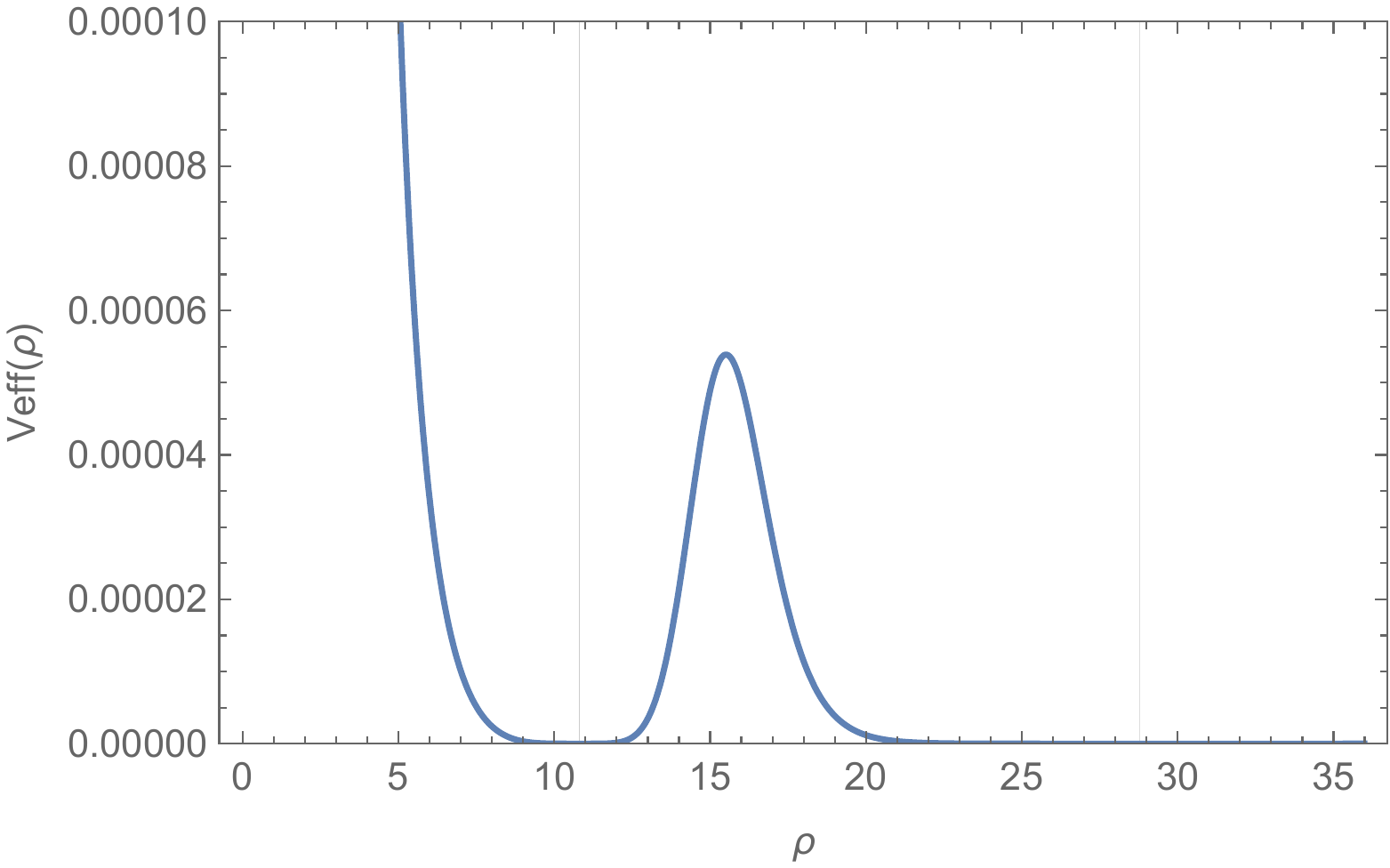}\\
\includegraphics[width=8cm]{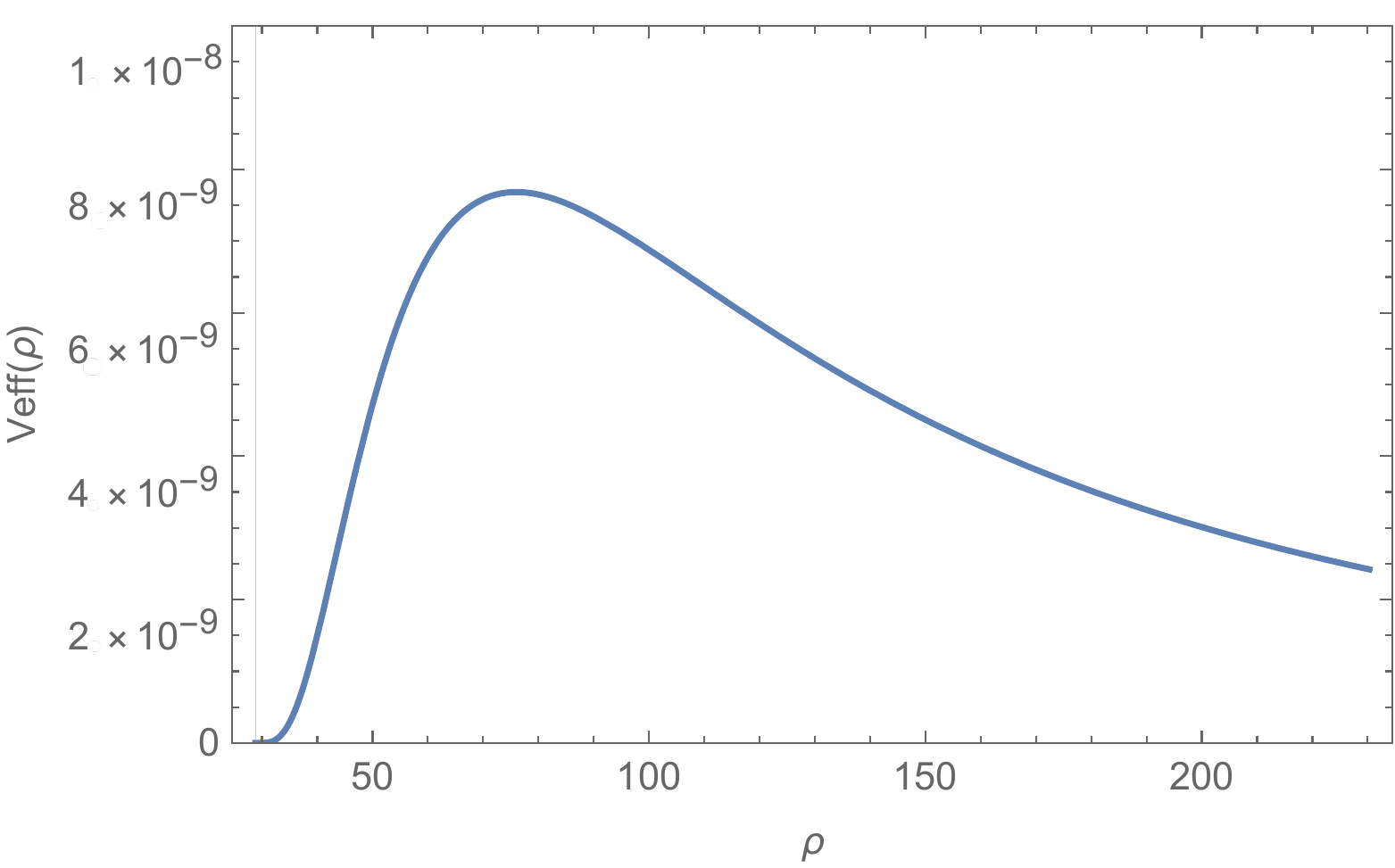}
\end{minipage}
\begin{minipage}{0.5\textwidth}
\centering
\includegraphics[width=6.5cm]{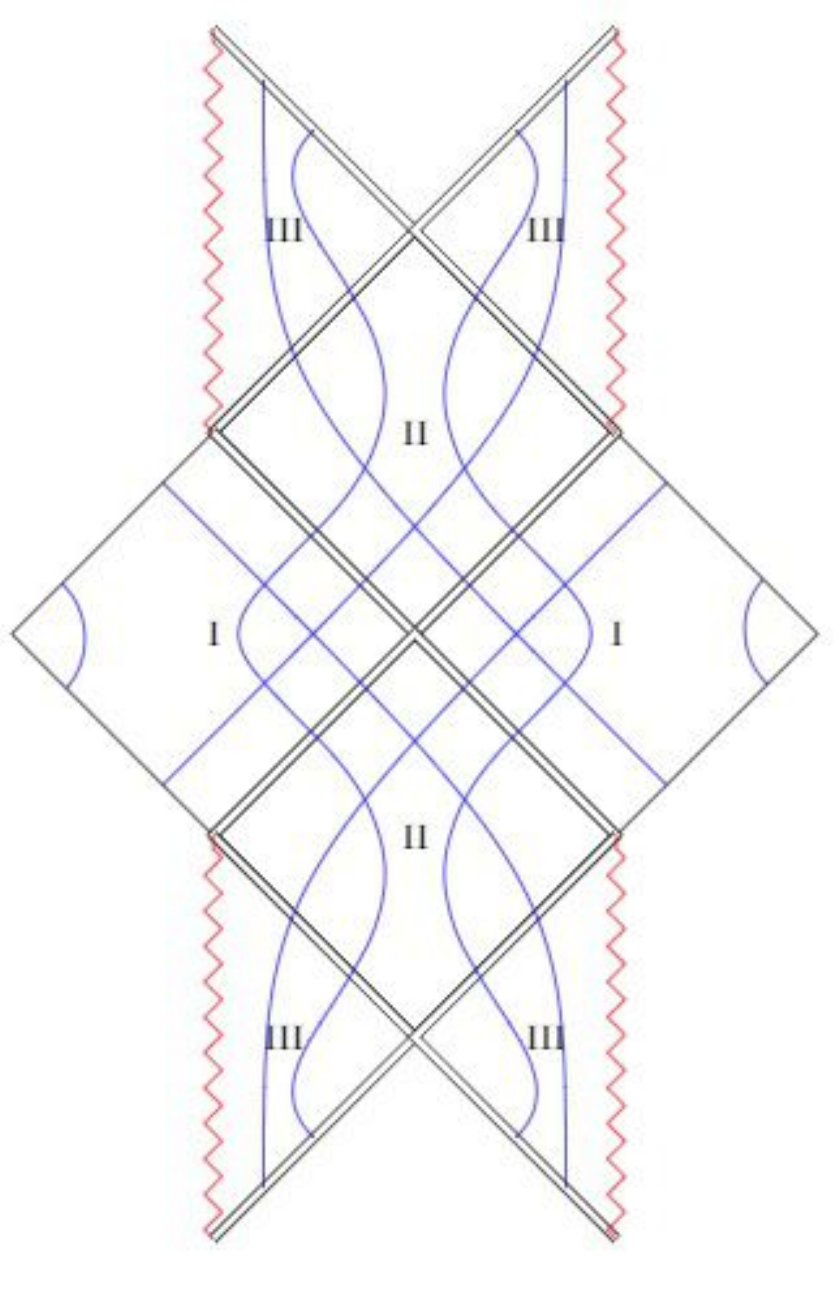}

\end{minipage}
\caption{\label{fig:veffsol1odd}Plot of the Effective potential for massless particles for the first solution, in~the case where $a$ assumes odd values (\textbf{left-top} figure), and~a zoom of the potential after the horizon (\textbf{left-bottom} figure). The~vertical lines in the left figure indicate the two event horizons of the solution. Moreover, the~right figure is the Carter--Penrose diagram drawing for this case. The~red lines are the singularities; the~double lines are the horizons; the blue lines are the possible geodesic curves. The~value of the parameters used were: $\lambda=27$, $k=9$, $m=30$, and $L=100$.}
\end{figure}

If $a$ assumes an odd value in the first solution of BHs, then we can directly take the $a$th root of Equation~(\ref{eq:sing}) to~obtain:
\begin{equation}
 1-\frac{2k}{\rho_s}=\left(\frac{m+\lambda}{m-\lambda}\right)^{1/a}\Rightarrow \rho_s=\frac{2k}{1-\left(\frac{m+\lambda}{m-\lambda}\right)^{1/a}} ,
\end{equation}
and since $m>\lambda$, the~singularity is located at $\rho=\rho_s < 0$. Therefore, in~this case, the~singularity is bounded by two event horizons, one at $\rho=0$ and another one at $\rho=2k$, where both surfaces have an infinite area, $e^{2\beta}\rightarrow\infty$, and~it is Minkowski at $\rho\rightarrow\infty$. Hence, we have three regions, called III, II, and I, respectively: $\rho_s<\rho<0$, $0<\rho<2k$, and $2k<\rho<\infty$. The~signature is $(+---)$ for Regions I and III, while it is $(-+--)$ for Region II. The~causal structure is similar to a non-extreme RN solution.
 In Figure~\ref{fig:veffsol1odd}, we plot the effective potential of the geodesics equation and the Carter--Penrose diagram (CPD) for this~case.

\subsubsection{Case of $a$ Even}
\label{seq:aeven}

\begin{figure}[H]
\centering
\begin{minipage}{0.5\textwidth}
\centering
\includegraphics[width=9cm]{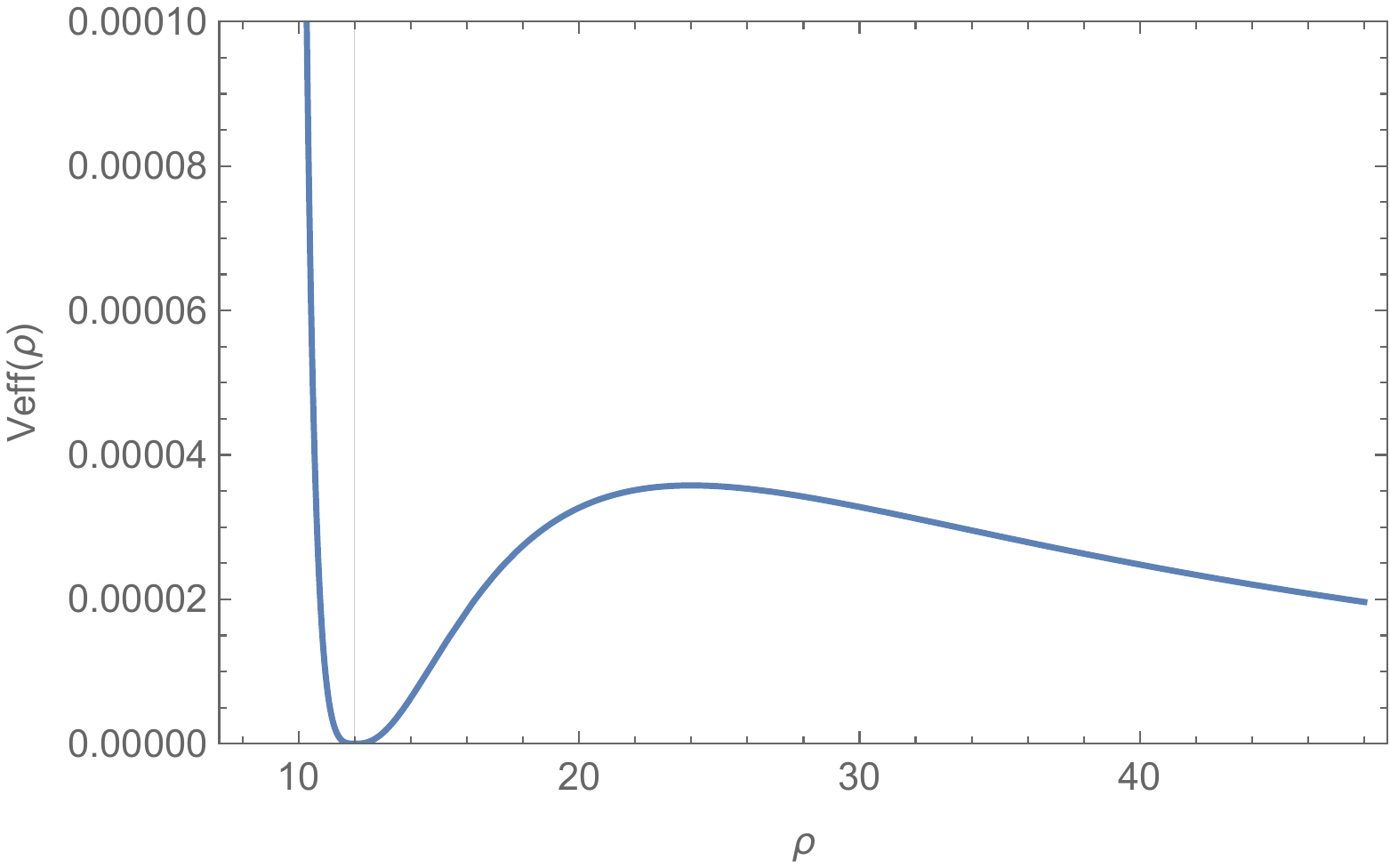}
\end{minipage}%
\begin{minipage}{0.5\textwidth}
\centering
\includegraphics[width=4cm]{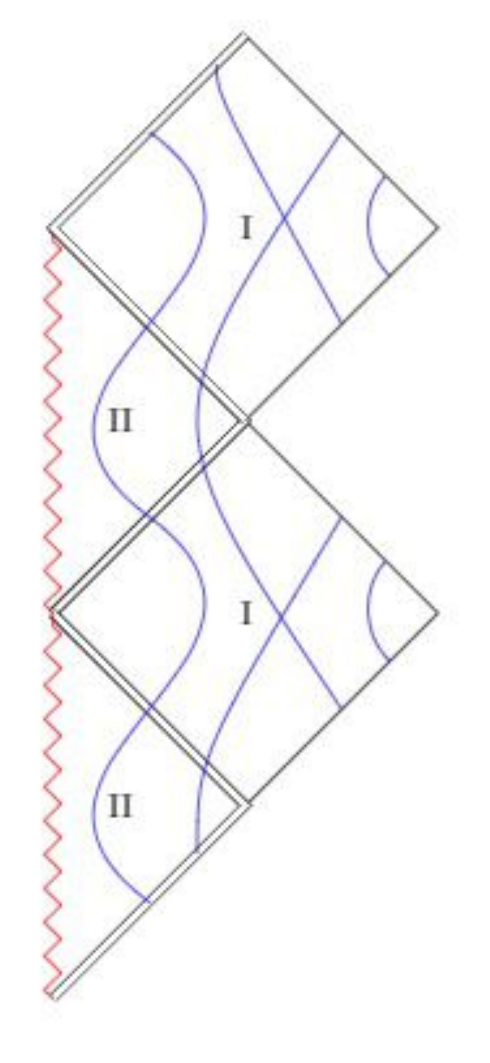}
\end{minipage}
\vspace{-9pt}
\caption{\label{fig:veffsol1even}Plot of the effective potential for massless particles (\textbf{left} figure) and the Carter--Penrose diagram (\textbf{right} figure) for the first solution with $a$ even. The~vertical line indicates the event horizons. In~the CPD, the red lines are the singularities, the~double lines are the horizons, and the blue lines are the possible geodesic curves. Here, we used the following values for the parameters: $\lambda=6$, $k=3$, $m=10$, $L=100$.}
\end{figure}

Now if $a$ is even in the first solution, then the~$a\mbox{th}$ root of Equation~(\ref{eq:sing}) gives us:
\begin{equation}
 1-\frac{2k}{\rho_s}=\pm\left(\frac{m+\lambda}{m-\lambda}\right)^{1/a}\rightarrow \rho_{s,\pm}=\frac{2k}{1\pm\left(\frac{m+\lambda}{m-\lambda}\right)^{1/a}},
\end{equation}
where the $\pm$ sign is due to the fact that $P(\rho_s)$ can be positive or negative. In~this sense, there are two singular surfaces, where one is the same as in the previous case with $a$ odd, $\rho_s=\rho_{s,-}$, while the other singular surface is located between the two horizons, $0<\rho_{s,+}<2k$. Analyzing the surface $\rho=\rho_{s,+}$, we find that the geodesics ends at this surface; thus, this solution consists of two regions, $\rho_{s,+}<\rho<2k$ and $2k<\rho<\infty$, instead of three. In~this case, there is no change of signature from one region to the other. It has a causal structure similar to the extreme RN solution. A~plot of the effective potential and the CPD for this case is shown in Figure~\ref{fig:veffsol1even}. 

\unskip

\subsection{Second Black Hole~Solution}
\label{seq:sec}
The second black hole solution in the phantom sector is {Equation}~(\ref{s-2}), and~the appropriate coordinate transformation for this case is:
\begin{equation}
 x=\frac{1}{u}.
\end{equation}

As a consequence, the~metric becomes,
\begin{equation}
\label{secsol}
 ds^2=h(x)dt^2-\frac{dx^2}{h(x)}-\frac{x^2}{h(x)}d\Omega^2,
\end{equation}
with:
\begin{equation}
 h(x)=\frac{4\lambda^2 \e^{-2\frac{\lambda}{x}}}{[(m+\lambda) - (m-\lambda)\e^{-2\frac{\lambda}{x}}]^2}.
\end{equation}

The geometric mass $m$ has the same definition as in the first BH solution, and~the constants are now related by:
\begin{equation}
\label{rel:sol2}
 m^2-Q^2=\lambda^2=\frac{C^2}{2}.
\end{equation}

The solution in this case has one parameter less than the first solution, since we fixed $k=0$; thus, the parameter $a$, which is a relation between $\lambda$ and $k$, is not defined here. We can also see that the transformation used to obtain {Equation}~(\ref{secsol}), giving an extension beyond the horizon, keeps the exponential nature of the functions in the metric; hence, there is no need for a ``quantization condition'' to extend the solution to the interior region. The~scalar and electric fields here are given by:
\begin{equation}
 \Phi(x)=\frac{C}{x} \quad \mbox{and}\quad E(x)=Qh(x).
\end{equation}

In this coordinate system, we can notice that this solution is Minkowski as $x\rightarrow\infty$, while a horizon is present at $x=0$. The~singularity is located at $x=x_s$, where $x_s$ satisfies the equation:
\begin{equation}
 \e^{-2\frac{\lambda}{x_s}}=\frac{m+\lambda}{m-\lambda}\Rightarrow x_s=\frac{-2\lambda}{\ln\left(\frac{m+\lambda}{m-\lambda}\right)}.
\end{equation}

\begin{figure}[H]
\centering
\begin{minipage}{0.5\textwidth}
\centering
\includegraphics[width=9cm]{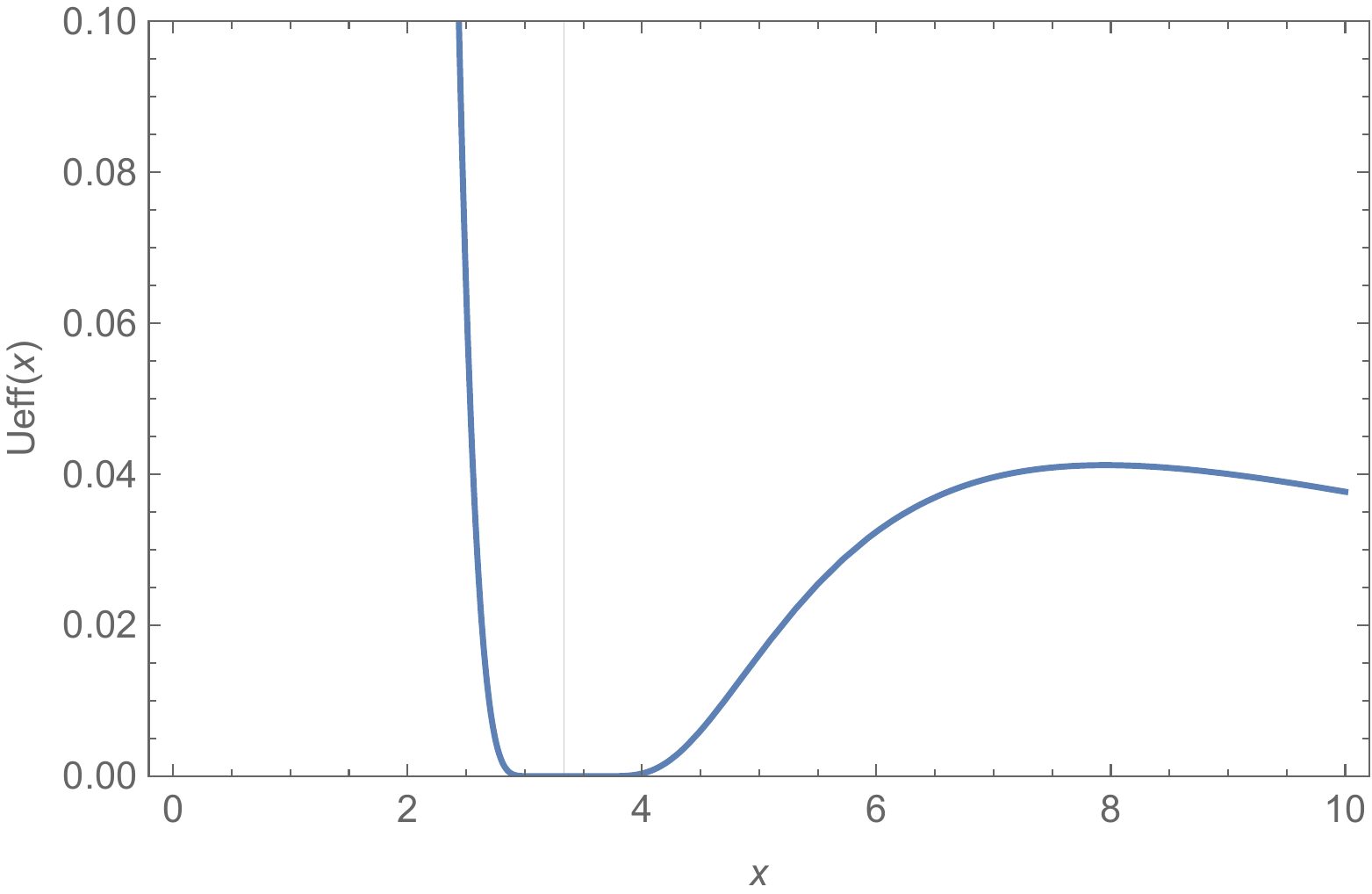}
\end{minipage}%
\begin{minipage}{0.5\textwidth}
\centering
\includegraphics[width=4cm]{cpdpar.pdf}
\end{minipage}
 \vspace{-9pt}
\caption{\label{fig:veffsol2}Plot of the effective potential for massless particles for the second solution (\textbf{left} figure). The~vertical line indicates the event horizon of the solution. Furthermore, the~right figure is the CPD for this solution, where the red lines are the singularities, the~double lines are the horizons, and the blue lines are the possible geodesic curves. For this plot, the~values for the parameters were: $\lambda=1.5$, $u_0=0.3$, and $L=10$.}
\end{figure}

 We can use the definition of the geometric mass, $m=\lambda\coth(\lambda u_0)$, to~write:
\begin{equation}
 x_s=-\frac{1}{u_0},
 \end{equation}
 and since $u_0>0$, we have that $x_s<0$. As~for the previous case, the~singularity here is timelike. Therefore, this solution, similar as in the first solution with $a$ even, has two regions, $x_s<x<0$ and $0<x<\infty$, connected by one event horizon at $x=0$. Figure~\ref{fig:veffsol2} shows a plot of the effective potential and the CPD of this~solution. 
 
 \vspace{-9pt}
\unskip

\subsection{\label{sec:wh}Wormhole~Solutions}

In the phantom sector ({Section}~\ref{seq:phantom}), we verified that there are three possible wormhole solutions. The~metric for these WHs can be written formally in a unified way by using the~transformation,
\begin{equation}
 u=\frac{1}{k}\arccot{|ky|},
\end{equation}
leading to,
\begin{equation}
\label{eq:gwh}
 ds^2=A_n(y)dt^2-\frac{dy^2}{A_n(y)}-{r_n(y)}^2d\Omega^2\quad\mbox{with}\quad {r_n(y)}^2=\frac{y^2+1}{A_n(y)},
\end{equation}
where the index $n$ corresponds to each wormhole solution.
The scalar and electric fields here are,
\begin{equation}
 \Phi(y)=\frac{C}{k}\arccot{ky}, \quad\mbox{and}\quad E(y)=QA_n(y).
\end{equation}

\begin{figure}[H]
\centering

\centering 

\begin{minipage}{0.32\textwidth}
\centering
\includegraphics[width=5.5cm]{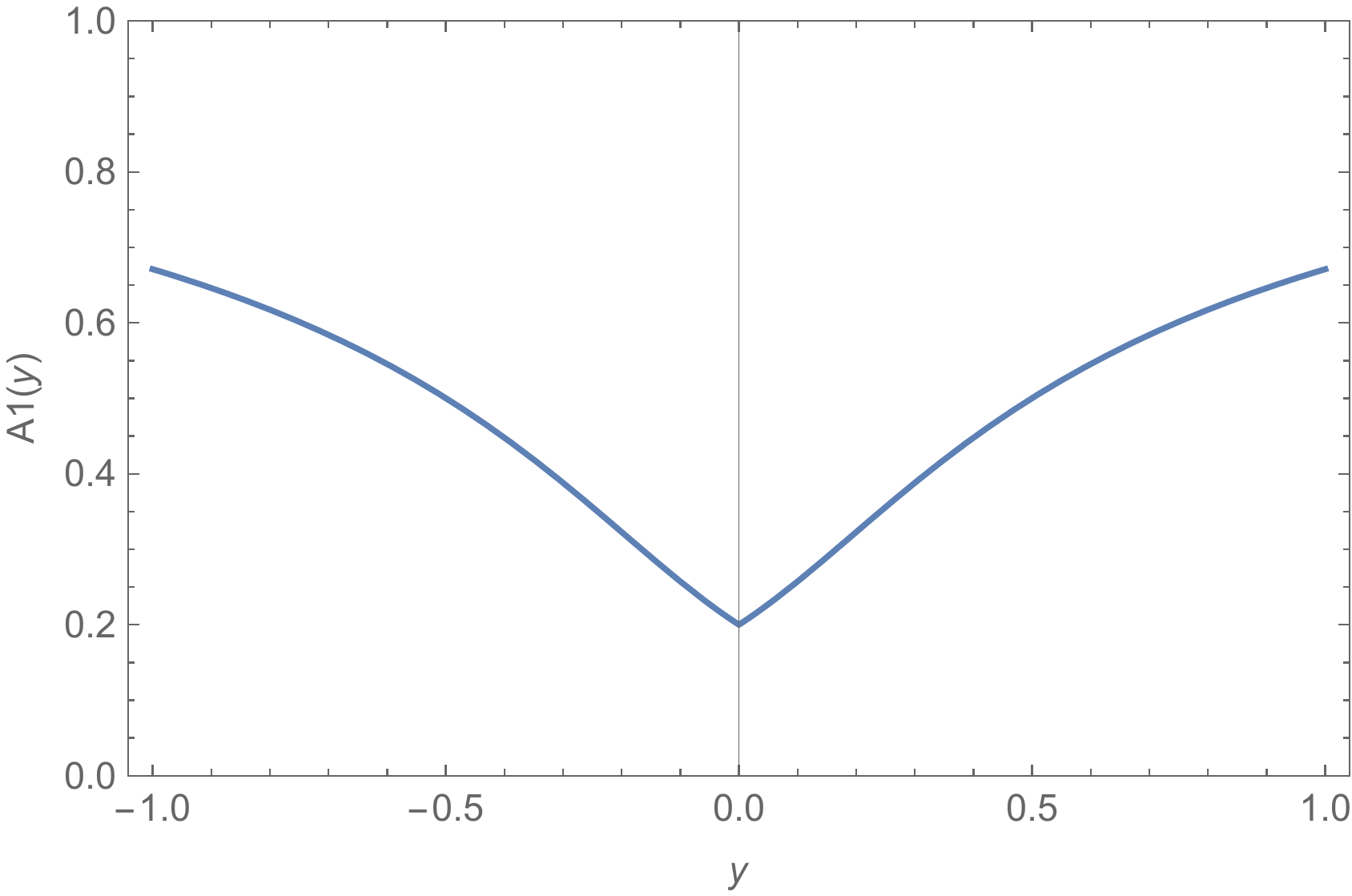}
\end{minipage}%
\begin{minipage}{0.32\textwidth}
\centering
\includegraphics[width=5.5cm]{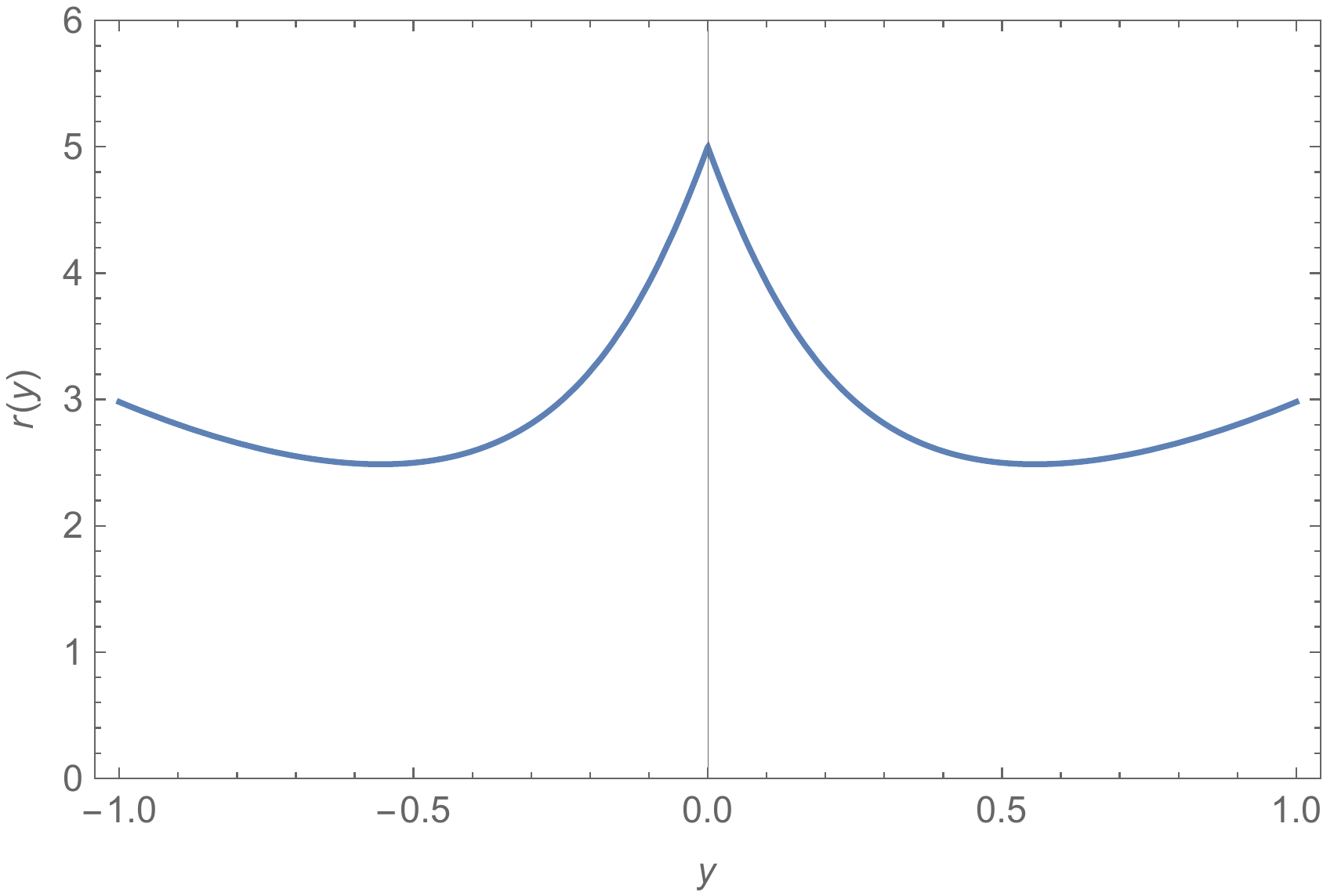}
\end{minipage}%
\begin{minipage}{0.32\textwidth}
\centering
\includegraphics[width=5.5cm]{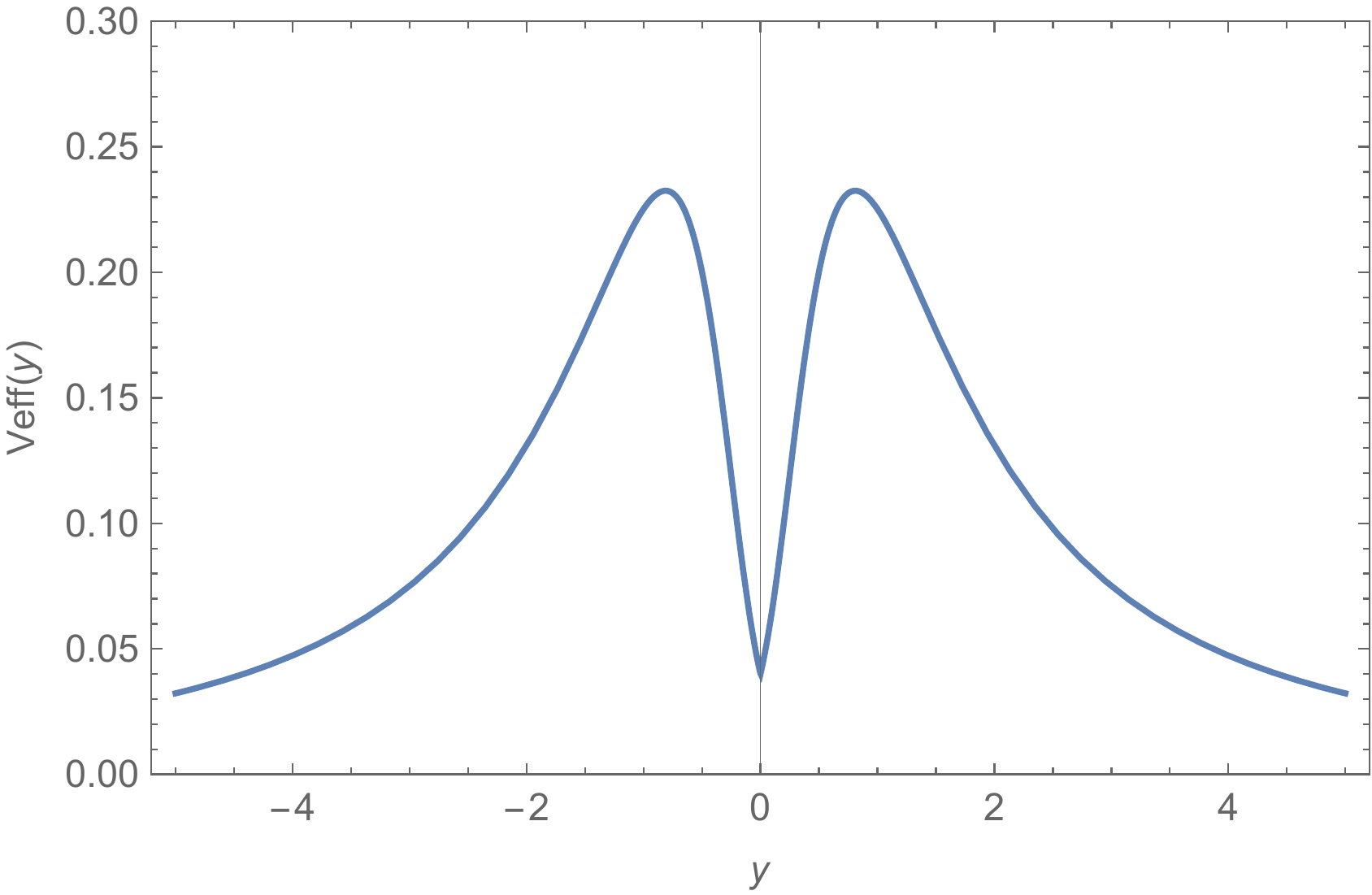}
\end{minipage}
\caption{\label{fig:wh1}Plot of the function $A_1(y)$, the~radius $r_1(y)$, and the effective potential, respectively, for~the first wormhole solution, {Equation}~(\ref{solwh1}). The~values used for the parameters in this plot were: $\lambda=1$, $k=3$, $m=2$, and~$L=1$.}
\end{figure}

\begin{figure}[H]
\vspace{-9pt}

\centering 

\begin{minipage}{0.32\textwidth}
\centering
\includegraphics[width=5.5cm]{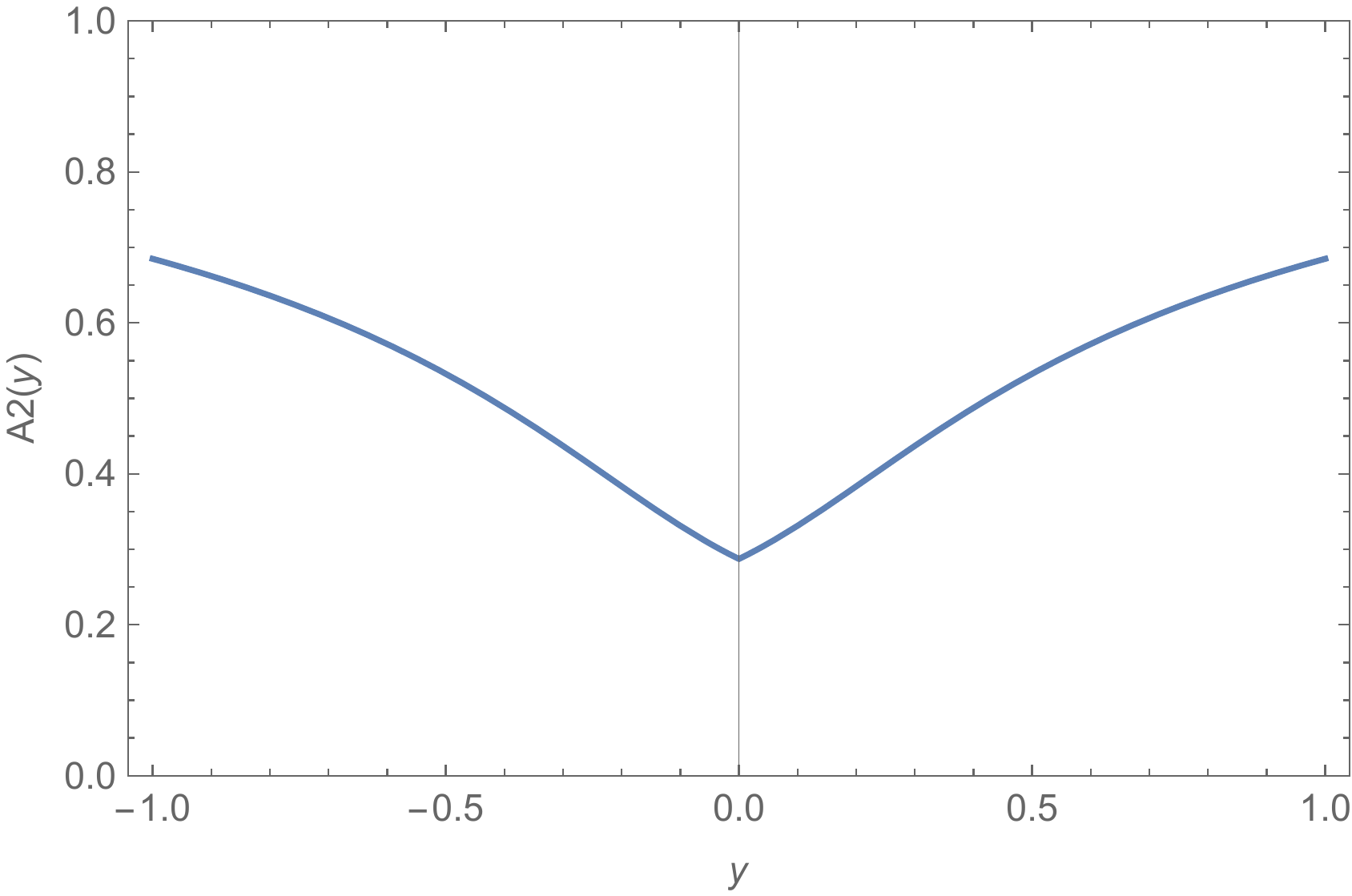}
\end{minipage}%
\begin{minipage}{0.32\textwidth}
\centering
\includegraphics[width=5.5cm]{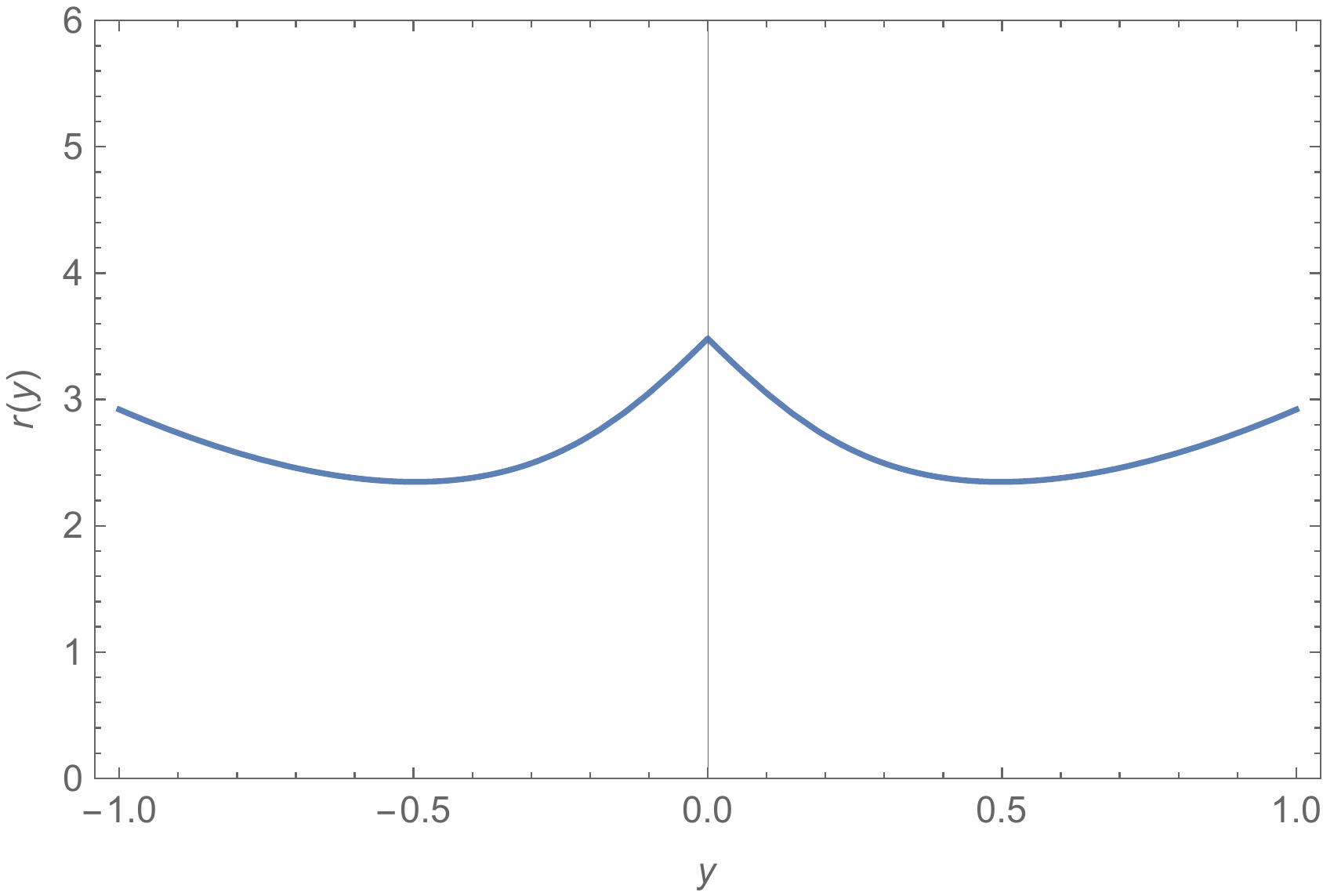}
\end{minipage}%
\begin{minipage}{0.32\textwidth}
\centering
\includegraphics[width=5.5cm]{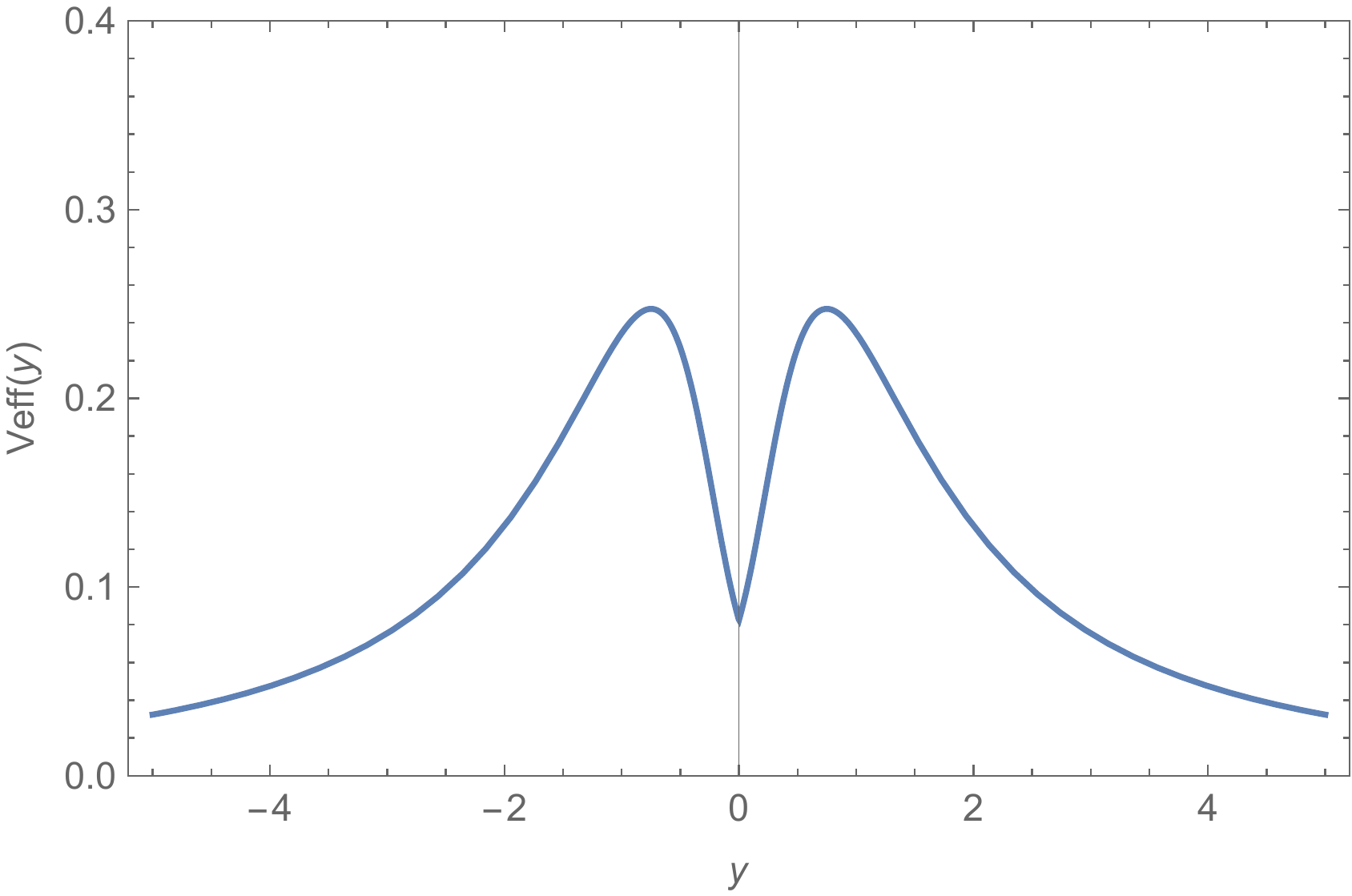}
\end{minipage}
\caption{\label{fig:wh2}Plot of the function $A_2(y)$, the~radius $r_2(y)$, and the effective potential, respectively, for~the second WH solution, {Equation}~(\ref{solwh2}). The~values used for the parameters in this plot were: $\lambda=1$, $k=3$, $m=2$, and~$L=1$.}
\end{figure}

\begin{figure}[H]
\centering
\vspace{-9pt}
\centering 
\begin{minipage}{0.32\textwidth}
\centering
\includegraphics[width=5.5cm]{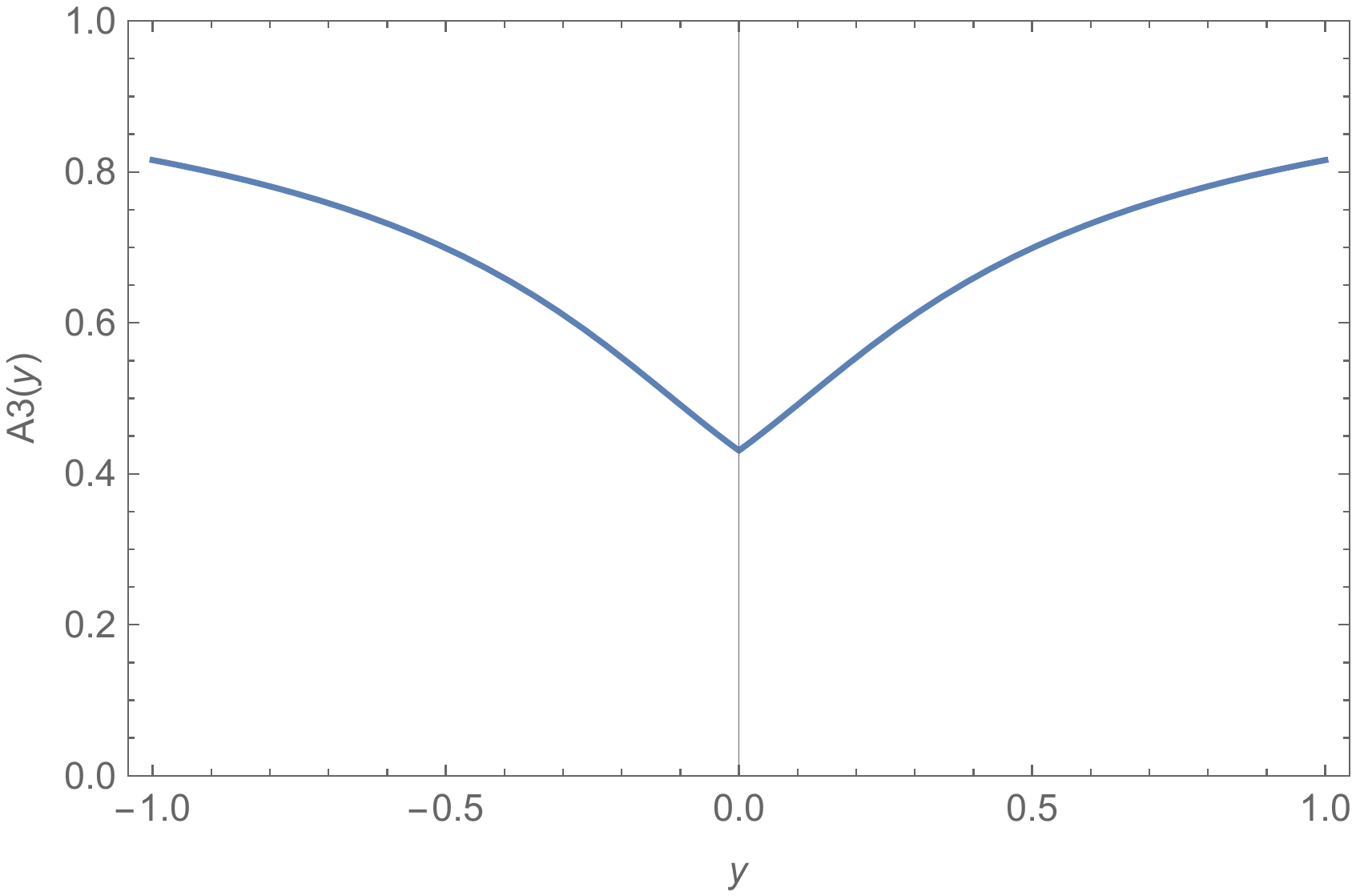}
\end{minipage}%
\begin{minipage}{0.32\textwidth}
\centering
\includegraphics[width=5.5cm]{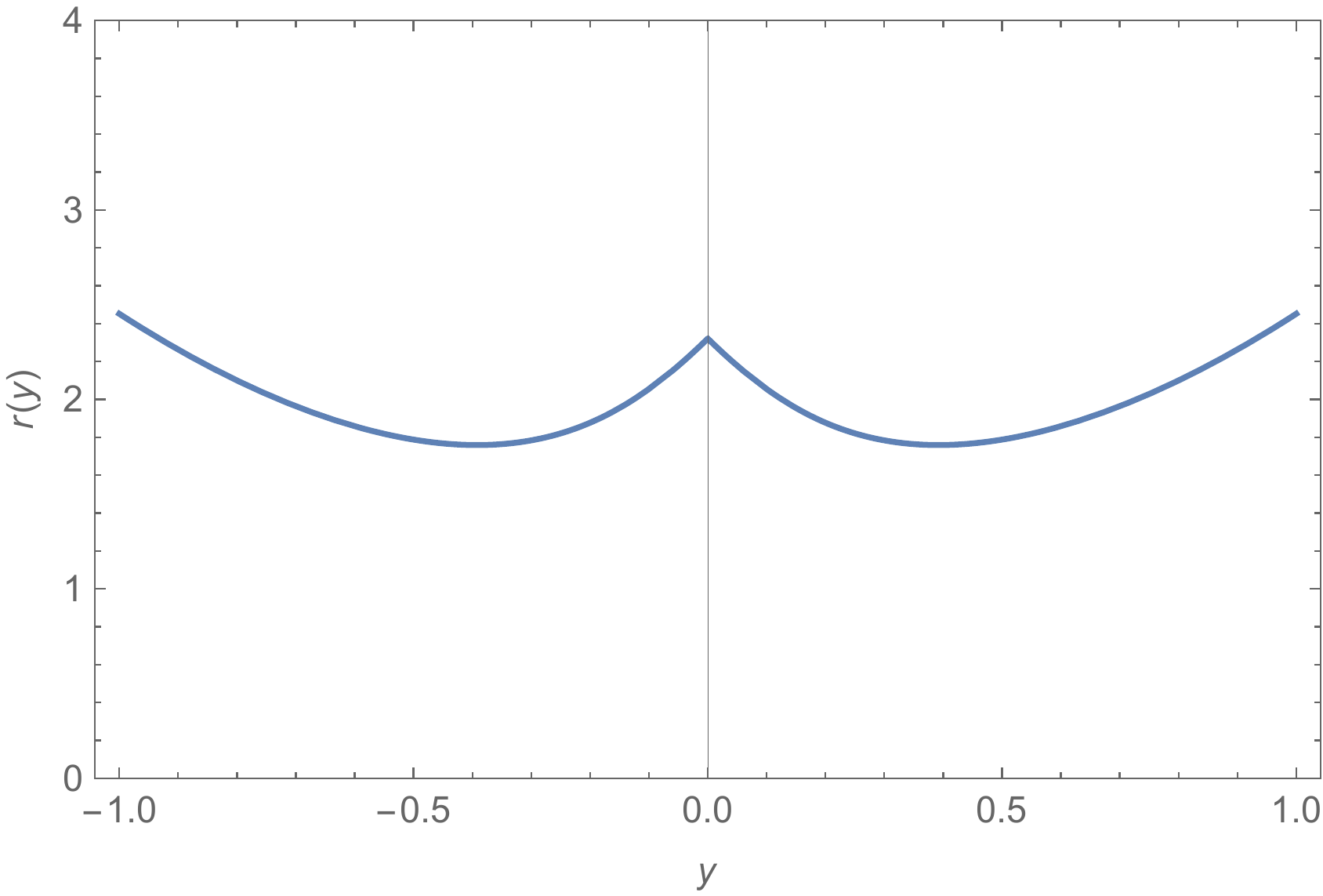}
\end{minipage}%
\begin{minipage}{0.32\textwidth}
\centering
\includegraphics[width=5.5cm]{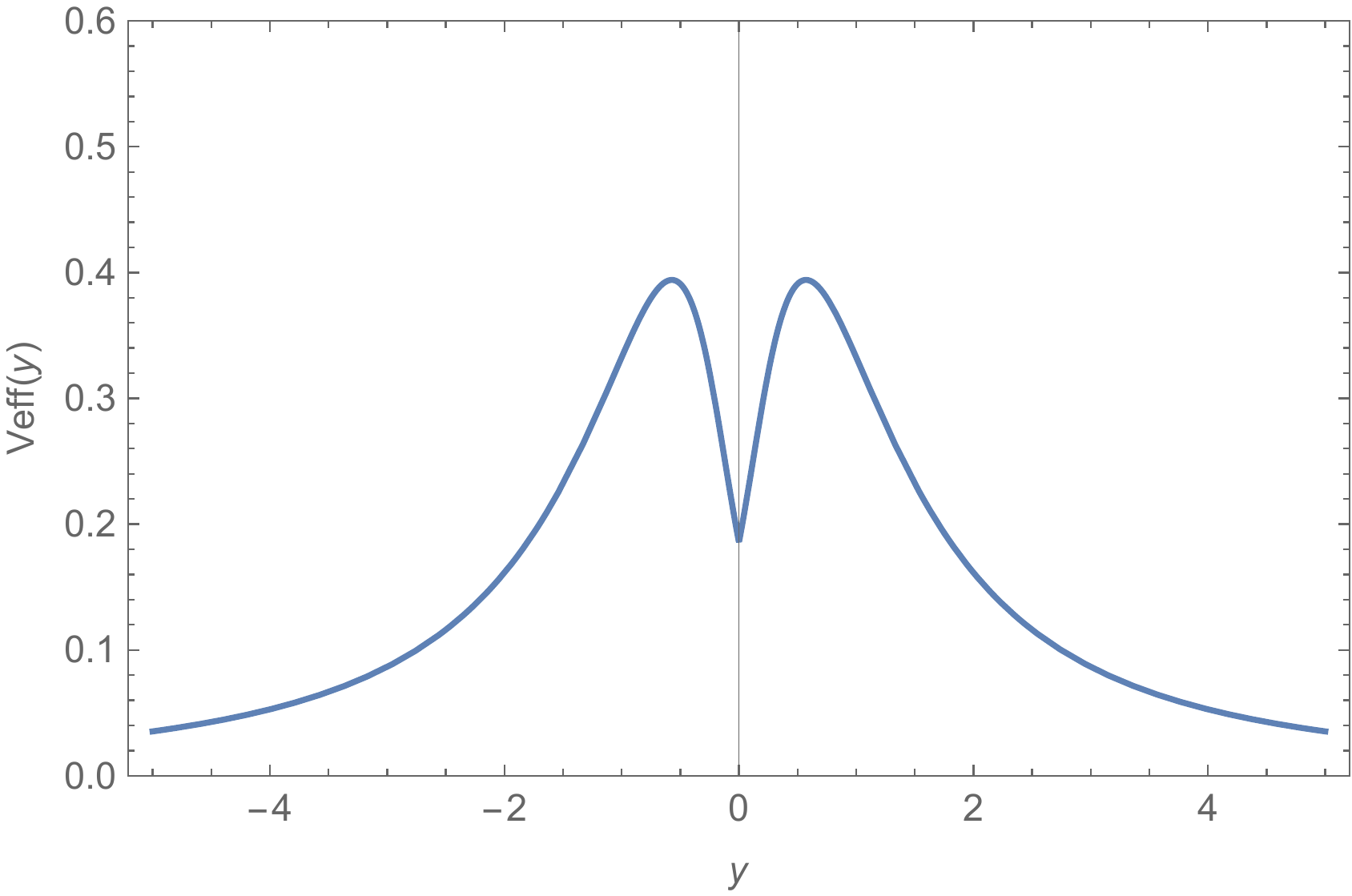}
\end{minipage}

\caption{\label{fig:wh3}Plot of the function $A_3(y)$, the~radius $r_3(y)$, and the effective potential, respectively, for~the third WH solution, {Equation}~(\ref{solwh3}). The~values used for the parameters in this plot were: $k=3$, $m=1$, and~$L=1$.}

\end{figure}

The metric (\ref{eq:gwh}) admits an analytical extension, so the range for the radial coordinate is $-\infty<y<\infty$, where $y\rightarrow\pm\infty$ corresponds to the two flat spatial infinities and the throat is at $y=0$, with~a radius of
$r_n(0)$. For~each of the three possible cases, the~function $A_n$ assumes a different~form.
\begin{itemize}
 \item First wormhole (\ref{solwhorm1}): In this case, we have the function:
\begin{equation}
\label{solwh1}
 A_1(y)=\frac{4\lambda^2\e^{-2\frac{\lambda}{k}\arccot{ky}}}{[(\lambda+m)+(\lambda-m)\e^{-2\frac{\lambda}{k}\arccot{ky}}]^2},
\end{equation}
 where:
\begin{equation}
  m^2-Q^2=\lambda^2=\frac{C^2}{2}-k^2.
 \end{equation}
 Since $\lambda$ is a real constant, the~above relation gives us $\lambda>m$, for~a non-null electric charge. The~radius of the throat in this case is:
\begin{equation}
  r_1(0)=\frac{(\lambda+m)+(\lambda-m)\e^{-2\frac{\lambda\pi}{2k}}}{2\lambda\e^{-\frac{\lambda\pi}{2k}}}.
 \end{equation}
 The effective potential of the geodesic motion for massless particles, has a local minimum at the throat and a maximum around it, where it is possible to have a photon sphere. In~Figure~\ref{fig:wh1}, there is a plot of $A_1$, the~radius $r_1$, and the effective potential. In~the limit $Q\rightarrow 0$, which corresponds to $\lambda\rightarrow m$, the~metric (\ref{solwh1}) takes the form of:
\begin{equation}
 ds^2=\e^{-2mu}dt^2-\e^{2mu}dy^2-\e^{2mu}(y^2+1)d\Omega^2.
\end{equation}
 The above metric is known as the wormhole of the anti-Fisher solution~\cite{bronnikov73}, and~for $m=0$, it becomes the Ellis wormhole~\cite{ellis}. A~more detailed discussion about these solutions can be found in~\cite{mex,bfz};

 \item Second wormhole (\ref{solwhorm2}): The function for this case has the form:
\begin{equation}
\label{solwh2}
 A_2(y)=\frac{\lambda^2{\sin(a\arccot{ky})}^{-2}}{[m+\lambda\cot(a \arccot{ky})]^2},
\end{equation}
where:
\begin{equation}
 Q^2-m^2=\lambda^2=k^2-\frac{C^2}{2}\quad\mbox{and}\quad a=\frac{\lambda}{k}.
\end{equation}

 Here, we have the condition that $|k|>|\lambda|$, which gives us that $a<1$. The~radius of the throat is:
\begin{equation}
 r_2(0)=\frac{m\sin(\frac{a\pi}{2})+\lambda\cos(\frac{a\pi}{2})}{\lambda}.
\end{equation}

The plot of the effective potential for geodesic motion for massless particles (Figure~\ref{fig:wh2}) shows that there is the possibility for photons to orbit the~throat;

 \item Third wormhole (\ref{solwhorm3}): Here, we have the function in the form:
\begin{equation}
\label{solwh3}
 A_3(y)=\frac{k^2}{[k+m\arccot{ky}]^2},
\end{equation}
where:
\begin{equation}
 Q^2-m^2=k^2-\frac{C^2}{2}=0.
\end{equation}

At the throat, $y=0$, the~radius is:
\begin{equation}
 r_3(0)=1+\frac{m\pi}{2k}.
\end{equation}

The effective potential in this case also allows photon spheres, as~we can see in Figure~\ref{fig:wh3}. The~metric in this case also becomes the Ellis wormhole~\cite{ellis} for $m=0$.

\end{itemize}

 \vspace{-3pt}

\section{Final Comments and~Conclusions} \label{sec:conclusions}

We analyzed, in~this work, the black hole and wormhole configurations of the Einstein--Maxwell system with a massless scalar field. Such a structure can be obtained from the Brans--Dicke theory in the presence of the Maxwell field by a conformal transformation. Wormholes' and black holes' charged solutions in the Brans--Dicke theory were studied in detail in~\cite{Bronnikov99}. In~the Brans--Dicke theory (Jordan frame) as in the Einstein--Maxwell-scalar system
(Einstein frame) studied here, black holes and wormholes can be obtained only when the scalar field is phantom, that is, it has negative kinetic~energy. 

The conformal transformation maps the seven main classes of solutions obtained in~\cite{Bronnikov99} (each of them containing sub-cases according to the conditions on the parameters) to all solutions obtained here. In~the present case, in~the Einstein frame, we have ten possible solutions, listed in Tables~\ref{table:3} and \ref{table:4}. In~all cases, there is a map directly to their counterparts in the Jordan frame. In~\cite{Bronnikov99},
 for~a canonical scalar field,
in both frames, only naked singularities are present. In~the phantom sector, the~possibilities are much richer, with~BH and WH configurations.
The possibilities are richer in the Jordan frame due to the presence of the conformal factor. For~example, for~$\lambda > k > 0$, in~the Jordan frame, there
are four possible cases, while in the Einstein frame, just two cases. Moreover, in~the Brans--Dicke original frame, the~central singularity can be timelike or spacelike, while
in the Einstein frame, both BH cases exhibit a timelike central singularity. The~WH solutions in one case are mapped into WH solutions in the other~frame. 

Although we can map all solutions obtained here to the ones obtained in the Jordan frame, these frames are not completely equivalent. The~presence of a conformal factor in the metric changes the geodesics followed by particles depending on the choice of the frames: a geodesic in the Jordan frame is not necessarily a geodesic in the Einstein frame
and vice versa. Even though, the~type of geodesic curves obtained here for the two BH solution, as~shown in Figures~\ref{fig:veffsol1odd}--\ref{fig:veffsol2}, are also present in the Jordan frame, which, however, contains in general more different types of curves than those found in the Einstein~frame.

The main properties of the black hole and wormhole solutions described here do not depend on the choice of the coordinate system, for~example the~fact that the black hole horizons have an infinite area. However, we must keep in mind that some coordinate systems may cover only a portion of the complete manifold, as happens if we fix the areal function as in the traditional Schwarzschild coordinates for which $e^{2\beta} = r^2$.

The different parameters of the model obey a ``quantization condition'', mainly due to the analytical extension beyond the horizon. These quantization conditions are different in the Jordan and Einstein frame due to the conformal factor. In the~Einstein frame, the~``quantization condition'' is given by {Equation}~(\ref{quantcond}), which leads to a different causal structure: one that is similar to the non-extreme Reissner--Nordstr\"om solution, the case of $a$ odd, while the other is similar to the extreme Reissner--Nordstr\"om solution, the case of $a$ even. In~both cases, the~central singularity is timelike, which seems to be a consequence of the presence of the electric field, as~in the Reissner--Nordstr\"om case. However, the~horizons, in~one case and in the other, have an infinite area, being an example of the so-called cold black holes. This may be considered as a consequence of the presence of a phantom scalar~field. 

The studies of the stability problem for the solutions found here are a necessary new step. For~an appropriate analysis of the problem, the use of the gauge-invariant quantities is required, as well as determining the stability of the model. From~the previous results obtained in the literature for similar configurations~\cite{mex,bfz,lv}, we may expect that the solutions are unstable. {We remark} that the BHs found here have some similarities with the RN black holes for which previous studies have revealed 
the presence of instabilities~\mbox{\cite{dotti1,dotti2}} mainly in the Cauchy horizon. However, the stability of objects such as black holes and wormholes must be studied case by case. We remark that the stability analysis implies writing a Schrödinger-type equation with an effective potential.
This effective potential is generally singular at the minimum of the areal function $e^{2\beta}$. This implies that a numerical analysis is necessary in order to complete the stability study. We hope to present such a numerical analysis in future~work.

{Finally}, it is important to give some words about the thermodynamics of the black hole solutions found here. They belong to the class of the so-called cold black holes because, in~principle, their Hawking temperature is zero. It is more appropriate to say that maybe it is not possible to attribute thermodynamics properties to such black holes, in~a sense similar to what happens with the extreme RN black holes; see~\cite{fabris} and~the references therein. Strictly speaking, we can only state that cold black holes have zero surface gravity $\kappa$, as~it can be verified by computing,
\begin{eqnarray}
\kappa = \frac{1}{2}\frac{g_{00}'}{\sqrt{g_{00}g_{11}}}\biggl|_{u_h}.
\end{eqnarray}

{It is} quite direct to verify that $\kappa$ is, for~the cases considered here, equal to zero. However, a~proper semiclassical evaluation of quantum fields in the spacetime of cold black holes reveals that all the computation is ill defined. Hence, it is not clear that it is possible to define the thermodynamic quantities as for the usual black holes. Especially, the~entropy of a black hole with zero temperature and/or an infinite surface horizon is not unambiguously defined; see~\cite{haw,zas}. To~our knowledge, the~thermodynamics of such objects remains an open problem.

\vspace{6pt}

\bigskip

{\bf Acknowledgements:} We thank CNPq (Brasil) and FAPES (Brasil) for partial financial support.

\newpage

\appendix

\section[\appendixname~\thesection]{Kretschmann~Scalar}
\label{sec:appendix}

When studying any spacetime, it is above all important to know whether it is regular, which means that all curvature invariants are finite at all its points, or~it contains curvature singularities at which at least one such
invariant is infinite. In~many cases, it is most helpful to check the finiteness
of the Kretschmann scalar, defined as:
\begin{equation}
  \mathcal{K}=R^{\mu\nu\lambda\sigma}R_{\mu\nu\lambda\sigma}.
 \end{equation}
 
For a spherically symmetric metric, the Kretschmann scalar is the sum of squares of all nonzero components of the Riemann
tensor~\cite{Bronnikovbook},
\begin{equation}
  \mathcal{K}=4(K_1)^2+8(K_2)^2+8(K_3)^2+4(K_4)^2,
 \end{equation}
where:
\begin{align}
  \label{kret:1}
  K_1&={R^{01}}_{01}=-\e^{-\alpha-\gamma}\left(\gamma'\e^{\gamma-\alpha}\right)',\\
  \label{kret:2}
  K_2&={R^{02}}_{02}={R^{03}}_{03}=-\e^{-2\alpha}\beta'\gamma',\\
  \label{kret:3}
  K_3&={R^{12}}_{12}={R^{13}}_{13}=-\e^{-\alpha-\beta}\left(\beta'\e^{\beta-\alpha}\right)',\\
  \label{kret:4}
  K_4&={R^{23}}_{23}=\e^{-2\beta}-\e^{-2\alpha}\beta'^2;
 \end{align}
 
 It is significant that all $K_i$ are invariant under the reparametrizations of the coordinate; in~other words, they behave as scalars at such transformations. Since the scalar $\mathcal{K}$ is a sum of squares, for~its finiteness, it is necessary and sufficient that all its components $K_i$ are~finite.
 
 It must be mentioned here that curvature singularities are not the only type of singularities that can appear in physically relevant spacetimes. In~the most general form, a~singularity is defined as a point or a set of points where geodesics terminate at a finite value of their affine parameter, characterizing geodesic~incompleteness.
 
\subsection{Kretschmann for the First Black Hole~Solution}
Using the first solution ({Equation}~(\ref{s-1})), one can find the scalars (\ref{kret:1}--\ref{kret:4}) as:

\begin{equation}
 K_1=-\frac{8\lambda^3\left[4\lambda(\lambda^2-m^2){P(\rho)}^{2a}-(m+\lambda)^2(k-\rho+\lambda){P(\rho)}^a+(m-\lambda)^2(k-\rho-\lambda){P(\rho)}^{3a}\right]}{\rho^4{P(\rho)}^2\left[(m+\lambda)-(m-\lambda){P(\rho)}^a\right]^4},
\end{equation}
\begin{equation}
 K_2=\frac{4\lambda^3\left[2\lambda(\lambda^2- m^2){P(\rho)}^{2a}-(m+\lambda)^2(k-\rho+\lambda){P(\rho)}^{a}+(m-\lambda)^2(k-\rho-\lambda){P(\rho)}^{3a}\right]}{\rho^4{P(\rho)}^2\left[(m+\lambda)-(m-\lambda){P(\rho)}^a\right]^4},
\end{equation}
\begin{equation}
 K_3=\frac{4\lambda^2\left[2(\lambda^2-m^2)(k^2-2\lambda^2){P(\rho)}^{2a}+(m+\lambda)^2((k-\rho)\lambda +k^2){P(\rho)}^a+(m-\lambda)^2((\rho-k)\lambda+k^2){P(\rho)}^{3a}\right]}{\rho^4{P(\rho)}^2\left[(m+\lambda)-(m-\lambda){P(\rho)}^a\right]^4},
\end{equation}
\begin{equation}
\small
 K_4=\frac{4\lambda^2\left[(2\lambda^2(k^2+m^2-\lambda^2)-2k^2m^2){P(\rho)}^{2a}+(m+\lambda)^2((\lambda+k)^2-2\lambda\rho ){P(\rho)}^a+(m-\lambda)^2((\lambda-k)^2+2\lambda\rho){P(\rho)}^{3a}\right]}{\rho^4{P(\rho)}^2\left[(m+\lambda)-(m-\lambda){P(\rho)}^a\right]^4}.
\end{equation}

\begin{figure}[H]
\centering 
\begin{minipage}{0.5\textwidth}
\centering
\includegraphics[width=8cm]{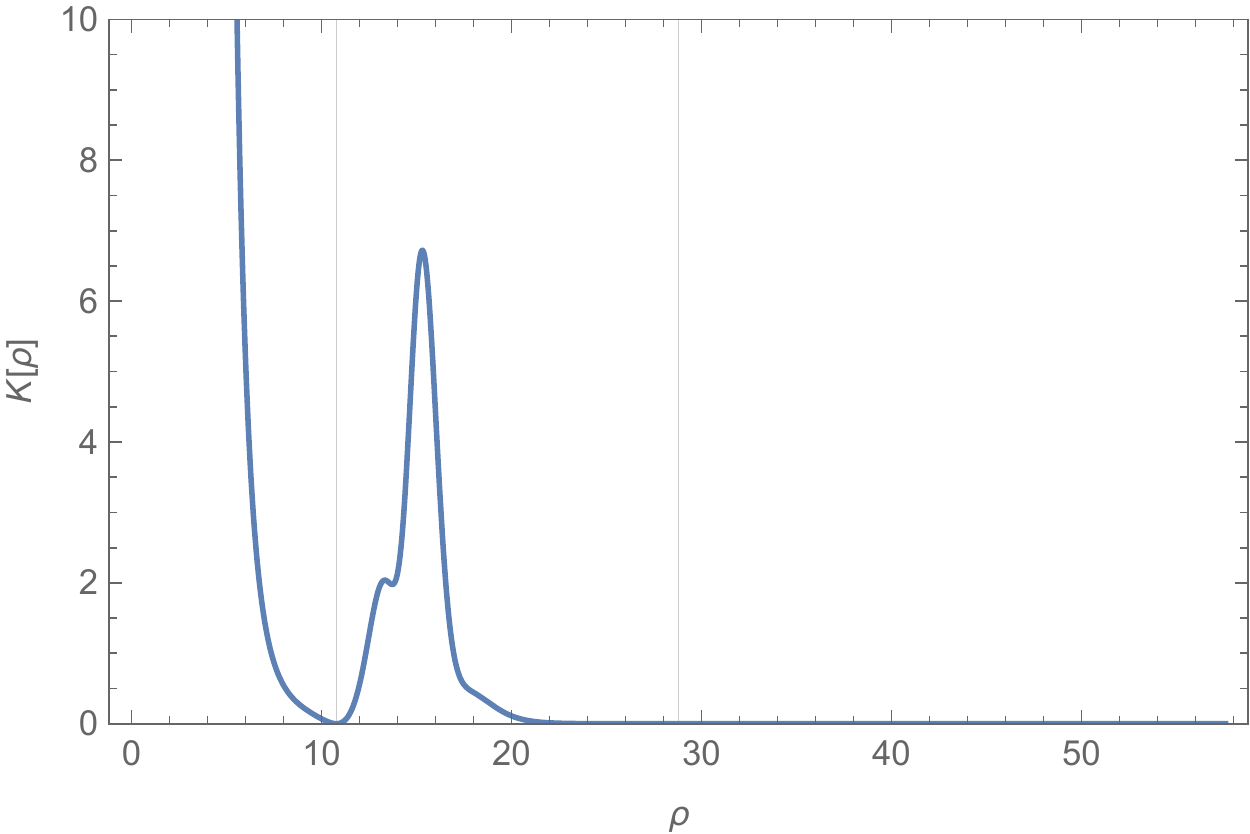}
\end{minipage}%
\begin{minipage}{0.5\textwidth}
\centering
\includegraphics[width=8cm]{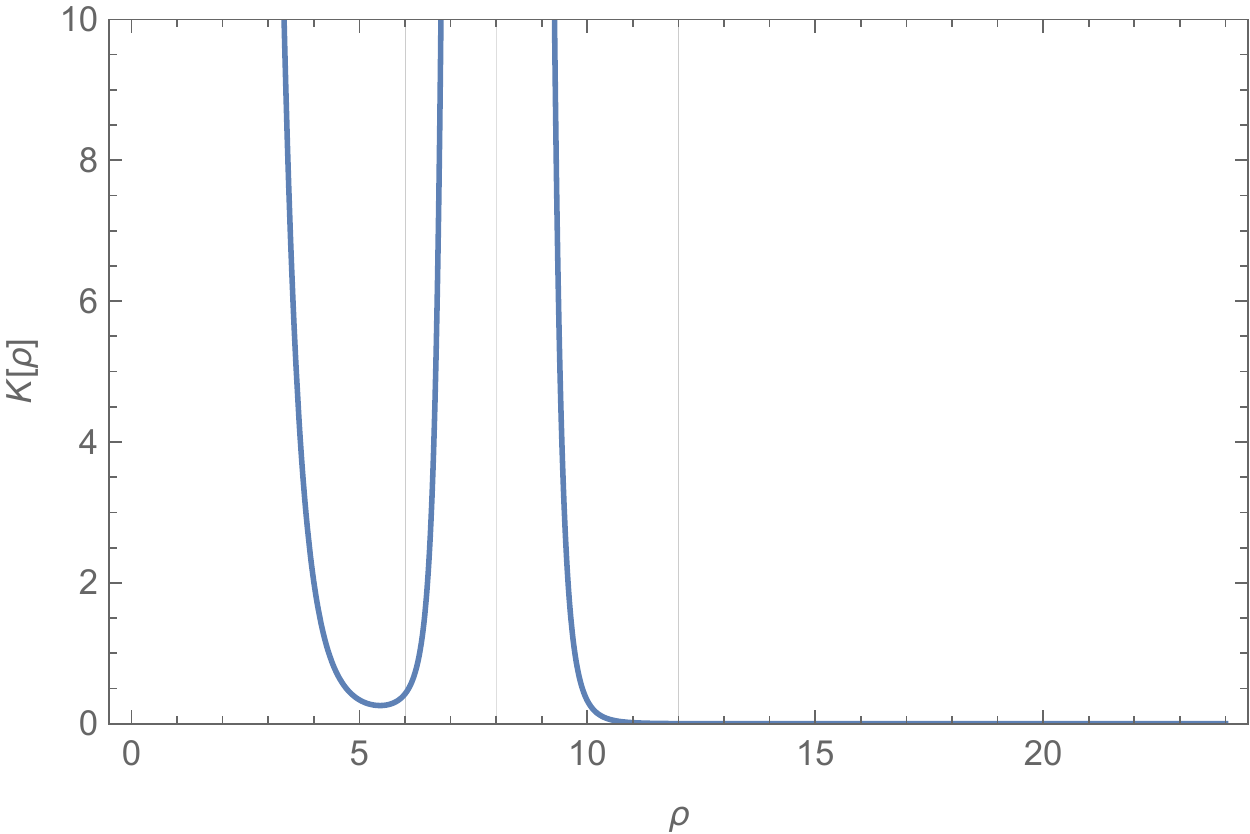}
\end{minipage}
\label{fig:kretsol1}
\caption{{Plot} 
 of the Kretschmann scalar for the first solution ({Equation}~(\ref{sol1})) with~$a$ odd (left figure) and with $a$ even (right figure). The~vertical lines in the left figure indicate the horizons, while in~the right figure, the~left and right vertical lines are the horizons and the one in the middle is the second singularity of the case of $a$ even, showing that the solution ends there, excluding the inner horizon. The~values of the parameter used here are the same as in the plots of {Figures}~\ref{fig:veffsol1odd} and~\ref{fig:veffsol1even}.}
\end{figure}
\unskip

\subsection{Kretschmann for the Second Black Hole~Solution}

For the second solution ({Equation}~(\ref{s-2})), the~scalars $K_i$ take the form:
\vspace{-6pt}

\begin{equation}
 K_1=\frac{-8\lambda^3\e^{\frac{2\lambda}{x}}\left\{(m+\lambda)^2(x-\lambda)\e^{\frac{4\lambda}{x}}-\left[4\lambda(m+\lambda) \e^{\frac{2\lambda}{x}}+(x+\lambda)(m-\lambda)\right](m-\lambda)\right\}}{x^4\left[(m+\lambda)\e^{\frac{2\lambda}{x}}-(m-\lambda)\right]^4},
\end{equation}
\begin{equation}
 K_2=\frac{4\lambda^3\e^{\frac{2\lambda}{x}}\left[(m+\lambda)\e^{\frac{2\lambda}{x}}+(m-\lambda)\right]\left[(x-\lambda)(m+\lambda)\e^{\frac{2\lambda}{x}}-(x+\lambda)(m-\lambda)\right]}{x^4\left[(m+\lambda)\e^{\frac{2\lambda}{x}}-(m-\lambda)\right]^4},
\end{equation}
\begin{equation}
 K_3=\frac{-4\lambda^3\e^{\frac{6\lambda}{x}}\left\{4\lambda(m^2-\lambda^2)\e^{-\frac{2\lambda}{x}}+\left[(m-\lambda)^2\e^{-\frac{4\lambda}{x}}-(m+\lambda)^2\right]x\right\}}{x^4\left[(m+\lambda)\e^{\frac{2\lambda}{x}}-(m-\lambda)\right]^4},
\end{equation}
\begin{equation}
 K_4=\frac{8\lambda^3\e^{\frac{6\lambda}{x}}\left\{(m-\lambda)^2\left(x+\frac{\lambda}{2}\right)\e^{-\frac{4\lambda}{x}}+(m+\lambda)\left[(m-\lambda)\lambda \e^{-\frac{2\lambda}{x}}-(m+\lambda)\left(x-\frac{\lambda}{2}\right)\right]\right\}}{x^4\left[(m+\lambda)\e^{\frac{2\lambda}{x}}-(m-\lambda)\right]^4},
\end{equation}

\begin{figure}[H]
\centering
\vspace{-9pt}
\includegraphics[scale=0.8]{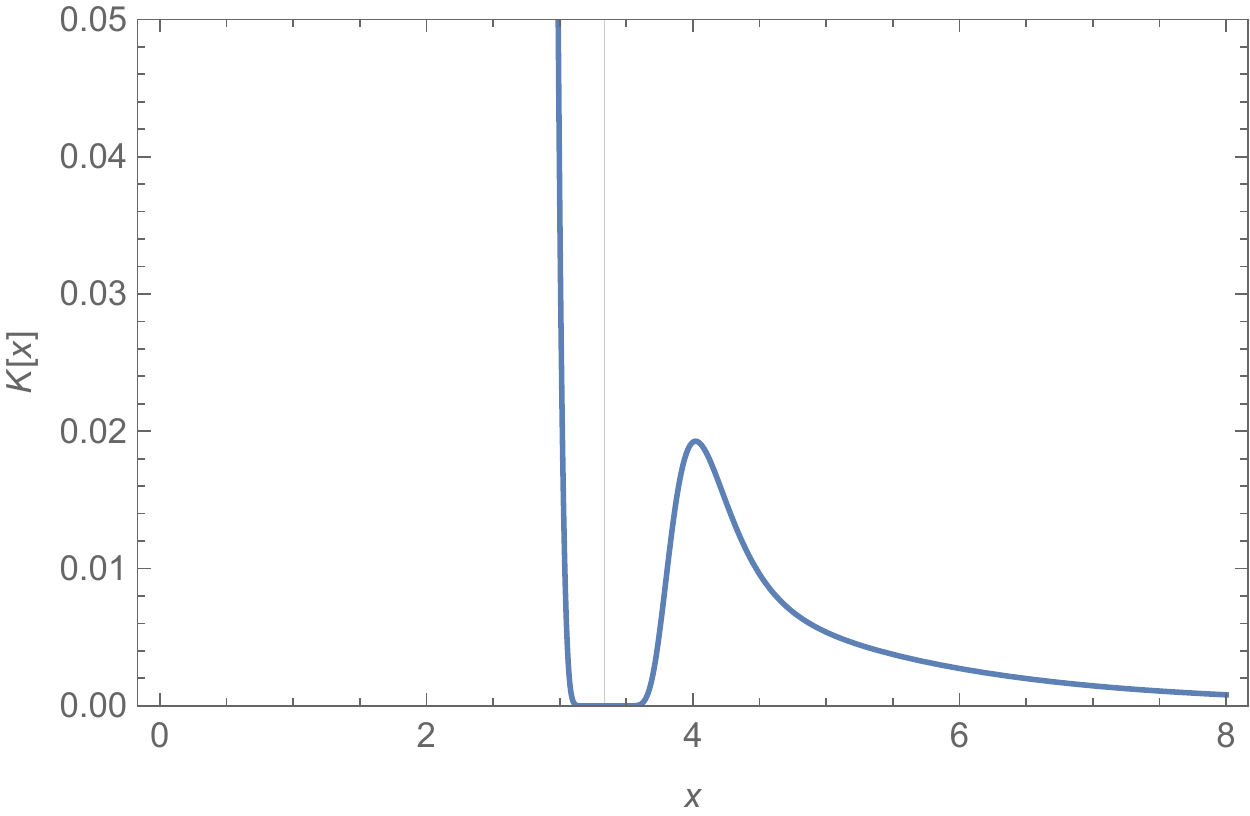}

\label{fig:kretsol2}
\caption{{Plot} of the Kretschmann scalar for the second solution {Equation}~(\ref{secsol}), where the vertical lines indicate the horizons. The~value of the parameters are the same as in Figure~\ref{fig:veffsol2}.}

\end{figure}
\unskip

\subsection{Kretschmann for Wormhole~Solutions}

The Kretschmann scalar for the WH solutions has several terms, which are not suitable to write here. However, we plot this scalar for each WH solution. The~plot shows us that the Kretschmann scalar, for~all cases, is finite~everywhere. 

\begin{figure}[H]
\centering
\begin{minipage}{0.33\textwidth}
\centering
\includegraphics[width=5.7cm]{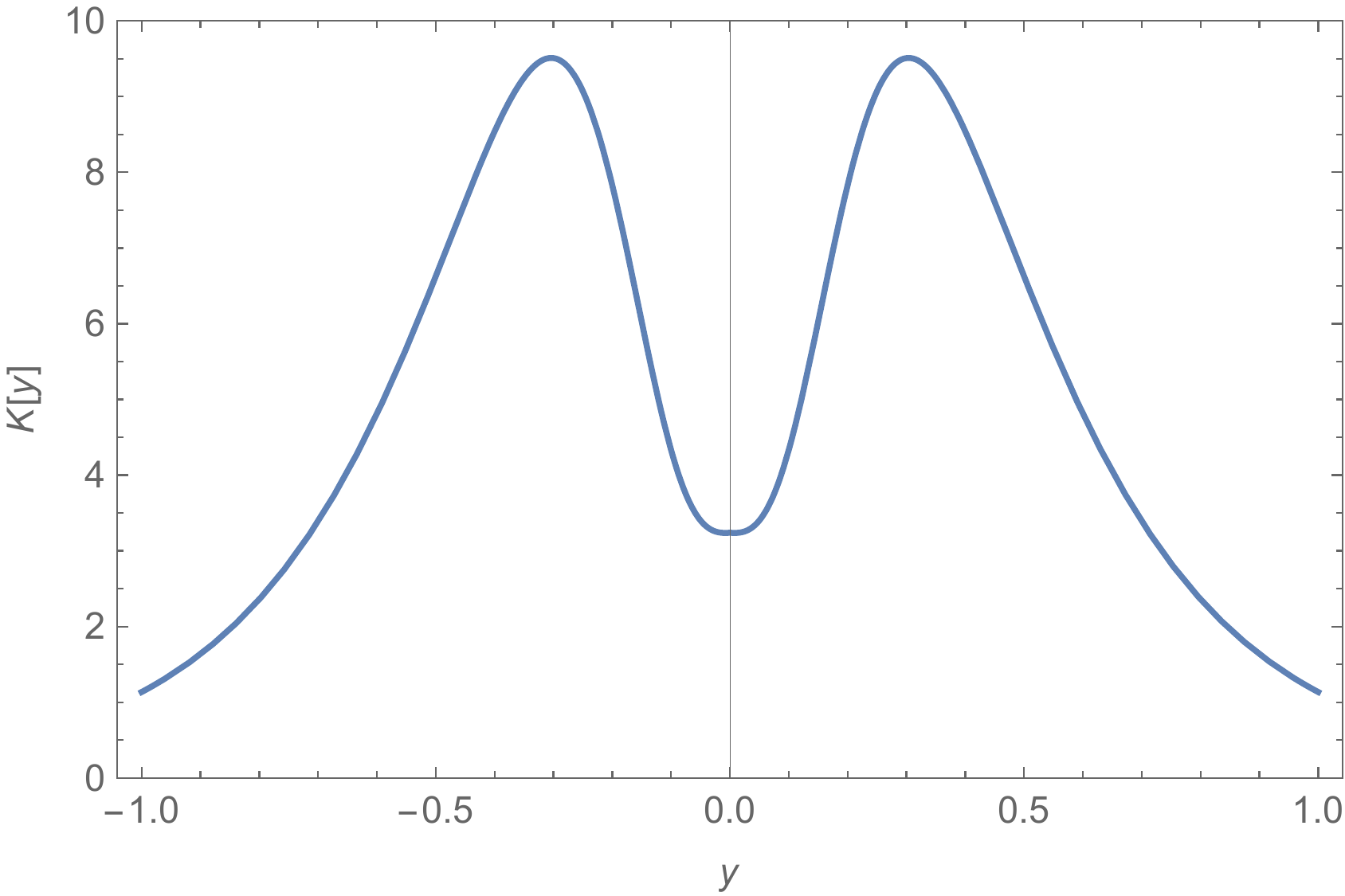}
\end{minipage}%
\begin{minipage}{0.33\textwidth}
\centering
\includegraphics[width=5.7cm]{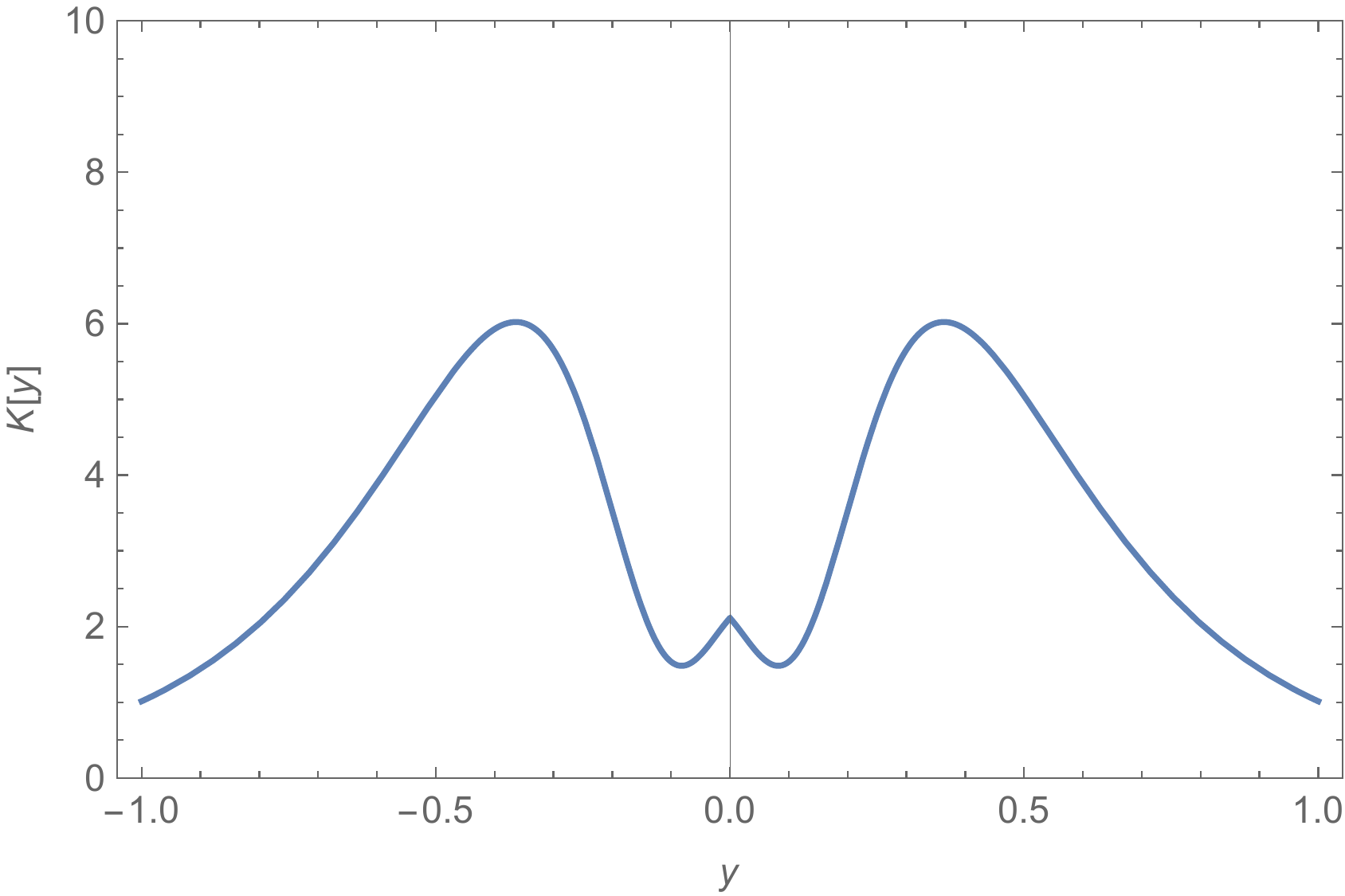}
\end{minipage}%
\begin{minipage}{0.33\textwidth}
\centering
\includegraphics[width=5.7cm]{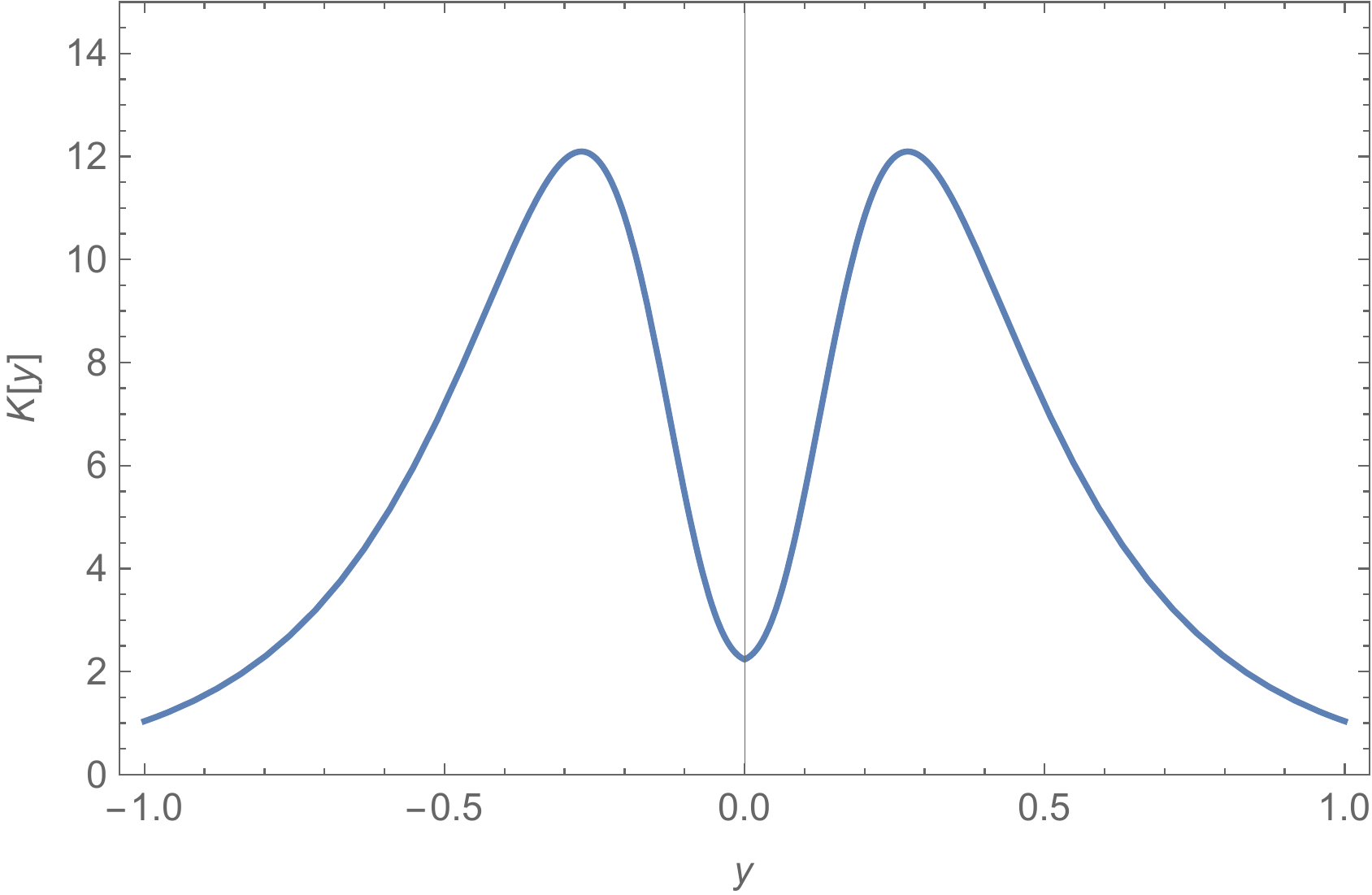}
\end{minipage}
\caption{{Plot} of the Kretschmann scalar for the first, second, and third WH solution ({Equations}~(\ref{solwh1}), (\ref{solwh2}) and (\ref{solwh3}), respectively). The~value of the parameters for each case is the same as in the plots of Figures~\ref{fig:wh1}--\ref{fig:wh3}, respectively.}
\end{figure}
\unskip

\section{General Form of Geodesic~Equations}

In this Appendix, we discuss the geodesic equations, using the Lagrangian formalism. The~Lagrangian of any spacetime can be written as:
\begin{equation}
 \mathcal{L}=\frac{1}{2}g^{\mu\nu}\dot{x}_\mu\dot{x}_\nu=\frac{1}{2}\mathcal{E},
\end{equation}
where:
\begin{equation}
 \mathcal{E}=\begin{cases}
 +&1 \quad\mbox{for timelike geodesics},\\
 &0 \quad\mbox{for null geodesics},\\
 -&1 \quad\mbox{for spacelike geodesics}.
 \end{cases}.
\end{equation}

For a static, spherically symmetric spacetime ({Equation}~(\ref{eq_01})), the~Lagrangian takes \mbox{the form:}
\begin{equation}
\label{lagrangian}
 \e^{2\gamma(u)}{\dot{t}}^2-\e^{2\alpha(u)}{\dot{u}}^2-\e^{2\beta(u)}\left[{\dot{\theta}}^2+\sin^2\theta{\dot{\phi}}^2\right]=\mathcal{E}.
\end{equation}

The symmetries of the system allow us to define the following conserved conjugate momentum,
\begin{equation}
 p_t=\frac{\partial\mathcal{L}}{\partial\dot{t}}=\e^{2\gamma}\dot{t}=E,
\end{equation}
\begin{equation}
 p_\phi=\frac{\partial\mathcal{L}}{\partial\dot{\phi}}=-\e^{2\beta}\sin^2\theta\dot{\phi}=L,
\end{equation}
{where $E$ and $L$ are the constants of integration. Substituting these expressions back into \mbox{{Equation}~(\ref{lagrangian})},}
\begin{equation}
 E^2e^{-2\gamma}-e^{2\alpha}{\dot{u}}^2-e^{2\beta}{\dot{\theta}}^2-\frac{L^2e^{-2\beta}}{\sin^2\theta}=\mathcal{E}.
\end{equation}

For simplicity, and~in connection with the symmetry of the problem, we assumed that the geodesic is located in the equatorial plane ($\theta=\frac{\pi}{2}$), leading to,
\begin{equation}
 e^{2\alpha+2\gamma}\dot{u}^2=E^2-\mathcal{E}e^{2\gamma}-L^2e^{2\gamma-2\beta}.
\end{equation}

\subsection*{Geodesic Equation for the General~Solution}

Now, for~the general solution ({Equation}~(\ref{gmetric})), the~geodesic equation becomes:
\begin{equation}
 \frac{\dot{u}^2}{{s(k,u)}^4}=E^2-\mathcal{E}\frac{{s(\lambda,u_0)}^2}{{s(\lambda,u+u_0)}^2}-L^2\frac{{s(\lambda,u_0)}^4{s(k,u)}^2}{{s(\lambda,u+u_0)}^4},
\end{equation}
where we can use the following transformation,
\begin{equation}
 \rho=\int\frac{du}{{s(k,u)}^2}\Rightarrow \dot{\rho}=\frac{\dot{u}}{{s(k,u)}^2},
\end{equation}
to obtain the geodesic equation in the form of an energy conservation law for a particle
moving in a potential field,
\begin{equation}
 \dot{\rho}^2=E^2-V_{eff}(u),
\end{equation}
with:
\begin{equation}
 V_{eff}(u)=\frac{{s(\lambda,u_0)}^2}{{s(\lambda,u+u_0)}^2}\left(\mathcal{E}+L^2\frac{{s(\lambda,u_0)}^2{s(k,u)}^2}{{s(\lambda,u+u_0)}^2}\right),
\end{equation}
playing the same role for geodesic motion as the potential in classical mechanics for a one-dimensional motion of a point particle: the motion is only possible in a region where $E^2 \geq V_{eff}(u)$, while the values of the coordinate at which $E^2 = V_{eff}(u)$ correspond to turning~points.

Remembering that $s(k,0)=0$, independent of the chosen sign of $k$, this describes two surfaces, $u=0$ and $u=-u_0$, the~asymptotic flat one and the singularity.
In the asymptotic region, the effective potential is a constant, while at the singularity, it diverges, $V_{eff}(-u_0)\rightarrow\infty$, which agrees with the expected behavior at these~surfaces.


\begin{thebibliography}{99}


\bibitem{herdeiro1} Herdeiro, C.A.R.; Radu, E. {Kerr black holes with scalar hair.} 
 \emph{Phys. Phys. Rev. Lett.} {\bf 2014}, \emph{112}, 221101. 

\medskip


\bibitem{japa1} Hong, J.P.; Suzuki, M.; Yamada, M. {Spherically symmetric scalar hair for charged black holes}. \emph{Phys. Rev.
Lett.} {\bf 2020}, \emph{125}, 111104.

\medskip

\bibitem{chandra} Chandrasekhar, S. {\em The Mathematical Theory of Black Holes}; Oxford University Press: Oxford, UK, 1992.

\medskip

\bibitem{visser} Visser, M. {\em Lorentzian Wormholes: From Einstein to Hawking}; AIP: Woodbury, NY, USA, 1996.

\medskip


\bibitem{frolov} Frolov, V.; Zelnikov, A. {\em Introduction to Black Hole Physics}; Oxford University Press: Oxford, UK, 2011.

\medskip

\bibitem{ligo}
 Abbott, B. P.; {Jawahar, S.; Lockerbie, N.A.; Tokmakov, K.V.} LIGO Scientific Collaboration and Virgo Collaboration.
 \emph{Phys. Rev. Lett.} {\bf 2016}, \emph{116}, 061102.

\medskip

\bibitem{eht}
 Event Horizon Telescope Collaboration.
 First M87 Event Horizon Telescope Results. I. The Shadow of the Supermassive Black Hole.
 \emph{Astrophys. J. Lett.} {\bf 2019}, \emph{875}, L1.

\medskip

\bibitem{herdeiro2} Herdeiro, C.A.R.; Paturyan, V.; Radu, E.; Tchrakian, D.H. {Reissner--Nordstr\"om black holes with non-Abelian hair}. \emph{Phys. Lett.} {\bf 2017}, \emph{772}, 63{--69}.

\medskip

\bibitem{Bronnikovbook}
 Bronnikov, K.A.; Rubin, S.G. \textit{Black Holes, Cosmology and Extra Dimensions}; World Scientific: Singapore, 2013.
 
\medskip

\bibitem{emd1} Gibbons, G.W.; Maeda, K. {Black holes and membranes in higher-dimensional theories with dilaton fields}. \emph{Nucl. Phys.} {\bf 1988}, \mbox{\emph{298}, 741{--775}.
}

\medskip

\bibitem{emd2} Garfinkle, D.; Horowitz, G.T.; Strominger, A. {Charged black holes in string theory.} \emph{Phys. Rev.} {\bf 1991}, \emph{43}, 3140{--3143}.
 
\medskip

\bibitem{emd3} Clément, G.; Leygnac, C. {Linear dilaton black holes.} \emph{Phys. Rev.} {\bf 2004}, \emph{70}, 084018.

\medskip

\bibitem{gmf} Clément, G.; Fabris, J.C.; Rodrigues, M.E. {Phantom black holes in Einstein--Maxwell-dilaton theory}. \emph{Phys. Rev.} {\bf 2009}, \emph{79}, 064021.

\medskip


\bibitem{Bronnikov99}
 Bronnikov, K.A.; Constantinidis, C.P.; Evangelista, R.L.; Fabris, J.C.
Electrically charged cold black holes in scalar--tensor theories. \emph{Int. J. Mod. Phys. } \textbf{1999}, \emph{8}, 481--505.
 
\medskip


\bibitem{bfz} 
 Bronnikov, K.A.; Fabris, J.C.; Zhidenko, A. {On the stability of scalar-vacuum spacetimes.} \emph{Eur. Phys J.} {\bf 2011}, \emph{71}, 1791.

\medskip


\bibitem{Bronnikovstr}
 Bronnikov, K.A.; Clément, G.; Constantinidis, C.P.; Fabris, J.C. 
 Structure and stability of cold scalar--tensor black holes. 
 \emph{Phys. Lett. A} \textbf{1998}, \emph{243}, 121--127. 

\medskip


\bibitem{Bronnikovsta}
 Bronnikov, K.A.; Clément, G.; Constantinidis, C.P.; Fabris, J.C. {Cold Scalar-Tensor Black Holes: Causal Structure, Geodesics, Stability.} 
 \emph{Grav. Cosmol.} {\bf 1998}, \emph{4}, 128--138.

\medskip


\bibitem{Bronnikovcoldbh}
 Bronnikov, K.A.; Fabris, J.C.; Pinto-Neto, N.; Rodrigues, M.E. {Cold black holes and conformal continuations.} \emph{Int. J. Mod. Phys.} {\bf 2008}, \emph{17}, 25.

\medskip


\bibitem{f(R)} De Felice, A.; {Tsujikawa, S.} {f(R) theories.} 
 \emph{Living Rev. Rel.} {\bf 2010}, \emph{13}, 3.

\medskip


\bibitem{bronnikov73}
 Bronnikov, K.A. {Scalar-tensor theory and scalar charge.}
 \textit{Acta Phys. Pol.} \textbf{1973}, \emph{4}, 251{--266}.

\medskip


\bibitem{wald} Wald, R.M. {\em General Relativity}; Chicago University Press: Chicago, IL, USA, 1984.

\medskip

 
\bibitem{ellis}
 {Ellis, H.G.} {Ether flow through a drainhole: A particle model in general relativity.}
 \emph{ J. Math. Phys.} \textbf{1973}, \emph{14}, 104{--118}.
 
\medskip


\bibitem{mex} 
 Gonzalez, J.A.; Guzman, F.S.; Sarbach, O. {Instability of wormholes supported by a ghost scalar field: I. Linear stability analysis.} \emph{Class. Quantum Grav.} {\bf 2009}, \emph{26}, 015010.
 
\medskip


 \bibitem{lv} Bronnikov, K.A.; Barcellos, V.A.G.; de Carvalho, L.P.; Fabris, J.C. {The simplest wormhole in Rastall and k-essence theories}. \emph{Eur. Phys. J.} {\bf 2021}, \emph{81}, 395.
 

\medskip

 
\bibitem{dotti1} Dotti, G.; Gleiser, R.J. {Gravitational instability of the inner static region of a Reissner-Nordstr\"{o}m black hole.} \emph{Class. Quantum Grav.} {\bf 2010}, \emph{27}, 185007.
 

\medskip

 
\bibitem{dotti2} Dotti, G. {Linear Stability of Black Holes and Naked Singularitie}. \emph{Universe} {\bf 2022}, \emph{8}, 38.
  
 
\medskip


\bibitem{fabris} Alvarenga, F.G.; Batista, A.B.; Fabris, J.C.; Marques, G.T. {Quantum modes around a scalar-tensor black hole: breakdown of the normalization conditions.} \emph{Grav. Cosm.} {\bf 2004}, \emph{10}, 184--186.
 
 
\medskip


\bibitem{haw} Hawking, S.W.; Horowitz, G.T.; Ross, S.F. {Entropy, area, and black hole pairs}. \emph{Phys. Rev.} {\bf 1995}, \emph{51}, 4302{--4314}.

\medskip


\bibitem{zas} Zaslavskii, O.B. {Thermodynamics of black holes with an infinite effective area of a horizon.} \emph{Class. Quantum Grav.} {\bf 2002}, \emph{19}, 3783{--3806}.

\end{thebibliography}
\end{document}